\documentclass[a4paper,fleqn,usenatbib,useAMS]{mnras}

\usepackage{graphicx}	
\usepackage{amsmath}	
\usepackage{amssymb}	
\usepackage{multicol}        
\usepackage{bm}		
\usepackage{pdflscape}	
\usepackage{breakurl}

\usepackage[T1]{fontenc}
\usepackage{ae,aecompl}

\usepackage{mathptmx}

\title[The transition from star formation to quiescence ] 
{SDSS IV MaNGA - the spatially resolved transition from star formation to quiescence.}

\author[F. Belfiore et al.] 
{Francesco Belfiore$^{1,2}\thanks{Email: fb338@cam.ac.uk}$,
Roberto Maiolino$^{1,2}$,
Claudia Maraston$^{3}$,
Eric Emsellem$^{4,5}$,
\newauthor 
Matthew A. Bershady$^{6}$, 
Karen L. Masters$^{3, 7}$, 
Dmitry Bizyaev$^{8,9}$,
M\'ed\'eric Boquien$^{10}$,
\newauthor 
Joel R. Brownstein$^{11}$, 
Kevin Bundy$^{12}$,
Aleksandar M. Diamond-Stanic$^{6, 13}$,
Niv Drory$^{14}$,
\newauthor 
Timothy M. Heckman$^{15}$,
David R. Law$^{16}$,
Olena Malanushenko$^{9}$,
Audrey Oravetz$^{9}$,
\newauthor 
Kaike Pan$^{9}$,
Alexandre Roman-Lopes$^{17}$,
Daniel Thomas$^{3}$, 
Anne-Marie Weijmans$^{18}$,
\newauthor 
Kyle B. Westfall$^{3}$ and
Renbin Yan$^{19}$
\\
\\
(Affiliations can be found after the references)
}

\pubyear{2015}

\begin{document}

\date{Accepted . Received ; in original form }
\pagerange{\pageref{firstpage}--\pageref{lastpage}} \pubyear{2016}

\maketitle
\label{firstpage}

\begin{abstract}
Using spatially resolved spectroscopy from SDSS-IV MaNGA we have demonstrated that low ionisation
emission line regions (LIERs) in local galaxies result from photoionisation by hot  evolved
stars, not active galactic nuclei, hence tracing galactic region hosting old stellar population where, despite
the presence of ionised gas, star formation is no longer occurring. LIERs are ubiquitous in both
quiescent galaxies and in the central regions of galaxies where star formation takes place at larger radii. We refer to these two classes of galaxies as extended LIER (eLIER) and central LIER (cLIER) galaxies respectively. 
cLIERs are late type galaxies primarily spread across the green valley, in the transition region between the star formation main sequence and quiescent galaxies. These galaxies display regular disc rotation in both stars and gas, although featuring a higher central stellar velocity dispersion than star forming galaxies of the same mass. cLIERs are consistent with being slowly quenched inside-out; the transformation is associated with massive bulges, pointing towards the importance of bulge growth via secular evolution.
eLIERs are morphologically early types and are indistinguishable from passive galaxies devoid of line emission in terms of their stellar populations, morphology and central stellar velocity dispersion. Ionised gas in eLIERs shows both disturbed and disc-like kinematics. When a large-scale flow/rotation is observed in the gas, it is often misaligned relative to the stellar component. These features indicate that eLIERs are passive galaxies harbouring a residual cold gas component, acquired mostly via external accretion. Importantly, quiescent galaxies devoid of line emission reside in denser environments and have significantly higher satellite fraction than eLIERs. Environmental effects thus represent the likely cause for the existence of line-less galaxies on the red sequence.

\end{abstract}



\section{Introduction} 
\label{intro}

Large spectroscopic galaxy surveys (SDSS, \citealt{York2000}, GAMA, \citealt{Driver2011}, zCOSMOS, \citealt{Lilly2007}) have demonstrated that galaxies are strongly bimodal in several of their fundamental properties, including colours, star formation rates, stellar ages, morphology and gas content \citep{Strateva2001, Blanton2003, Baldry2004, Baldry2006, Wyder2007, Schawinski2007, Blanton2009, Peng2010, Thomas2010}. These observations support a scenario where some physical processes lead galaxies to transition from the `star forming' (SF) blue cloud to the passive `red sequence', causing the eventual shut-down (or `quenching') of star formation on a galaxy-wide scale.

The physics driving this transformation is currently hotly debated and has not yet been unambiguously identified, although it is likely that more
than one mechanism is at play \citep{Schawinski2007, Peng2010, Schawinski2014, Mendel2013, Peng2015, Smethurst2015}. A growing body of evidence points towards at
least two quenching pathways: an \textit{environment-dependent} channel, acting differentially on central and satellite galaxies, and a
\textit{mass-dependent} channel, which may be associated with internal processes \citep{Silk1977, Rees1977,Kauffmann2006}. This inference is supported by the separability of the effects of mass and environment on the passive fraction \citep{Baldry2006, Peng2010, Thomas2010}. These two quenching pathways have each been associated with several physical processes: ram pressure stripping \citep{Larson1980}, harassment and strangulation \citep{Gunn1972, Peng2015} for the environment-related channel and radio mode feedback from active galactic nuclei (AGN, \citealt{Dekel2006, Croton2006, Cattaneo2009}), halo mass shock heating \citep{Birnboim2003, Keres2005, Dekel2006, Keres2009} or morphological quenching \citep{Martig2009, Genzel2014, Forbes2014} for the mass-dependant channel. It has also been suggested that both mass and environmental quenching may be manifestations of the same quenching channel \citep{Knobel2015, Carollo2016}.

Observationally much work has been dedicated to the study of environmental effects, demonstrating that local environment correlates well with morphology \citep[e.g][]{Dressler1980, Bamford2009}, colour \citep{Blanton2005a, Baldry2006, Peng2010} or gas content \citep{Cortese2011, Boselli2014}. The direct effect of the hot cluster atmosphere on infalling galaxies (ram pressure stripping) is further observed both via trails of H\textsc{i} \citep[e.g][]{Chung2009} and ionised gas \citep[e.g][]{Fumagalli2014, Poggianti2016}. 

Regarding the mass quenching regime, the observational picture is much more complex. Several studies agree on the fact that the presence of a massive bulge is the single best observational parameter correlating with the probability of a galaxy to be passive \citep{Pasquali2012, Cheung2012, Wake2012, Bluck2014}. Since black hole mass is correlated to the bulge mass \citep{Marconi2003, Haring2004, McConnell2013}, it can be expected that galaxies with more massive bulges experience a larger amount of energetic feedback from their black holes. The details of the coupling between the black hole and its host galaxy are, however, not currently understood. As recently reviewed by \cite{Heckman2014}, AGN appear as a bimodal population that can be broadly divided into radiative-mode (including Seyfert galaxies and quasars) and radio-mode AGN. In the low redshift Universe, no direct correlation is observed between radiative-mode AGN and the shutdown of star formation,
as radiative-mode AGN are often associated with the presence of star formation in the central regions of galaxies (inner few kpc). While
radiative-mode AGN are observed to power large scale outflows \citep{Maiolino2012, Cicone2014, Carniani2015}, the duty cycle and time-scales of
these phenomena is uncertain, thus making their impact on the exhaustion of the cold gas reservoir in galaxies questionable. Radio mode AGN, on
the other hand, are observed to inflate large bubbles of ionised gas in massive haloes, thus playing an important contribution to offsetting the
cooling of the gas in the halo \citep{Fabian2012}. The extension of this feedback mechanisms to lower masses (halo mass $< 10^{12}~M_\odot$) is observationally uncertain.

Both environmental and mass quenching (through radio-mode AGN heating) channels are included in recent semi-analytical models \citep{Croton2006, Bower2006} and in cosmological hydrodynamical simulations via sub-grid prescriptions \citep{Gabor2012, Vogelsberger2014, Schaye2015} in order to match the stellar mass function and the bimodality of the observed galaxy population. In recent years, models of galaxy formation and evolution have become able to qualitatively reproduce the main features of the bimodal galaxy population \citep[e.g.][]{Sales2015, Trayford2016}. Namely, in simulations environmental quenching appears to be the main cause of growth of the red sequence at low masses, while AGN radio-mode heating is responsible for the presence of high-mass central passive galaxies. However, while qualitative agreement with observations is promising, this success does not imply that the sub-grid recipes implemented for AGN feedback are physically correct. 

Observationally, further insight into the quenching process could be obtained by a statistical study of how it proceeds inside galaxies: `inside
out' versus `outside in' scenarios \citep[e.g][]{Tacchella2015, Li2015}, and what role the different morphological components (disc, bar, bulge) have
in the process \citep{Martig2009, Masters2012}. Current integral field spectroscopy (IFS) galaxy surveys (CALIFA, \citealt{Sanchez2012a}; SAMI,
\citealt{Allen2015}; MaNGA, \citealt{Bundy2015}) promise to revolutionise the observational picture, by enabling the study of the relation
between star formation and quiescence on resolved scales. Nebular emission lines, originating from SF regions, are
the ideal tool to trace current ($< $100 Myr old) star formation on the kpc scales \citep{Kennicutt1998} probed by a survey like MaNGA (mean redshift $<z> \sim 0.03$), provided other contributions to line emission (AGN, old hot stars etc) can be disentangled.

In this paper we make use of a large sample of $\sim$ 600 galaxies observed as part of the Sloan digital sky survey IV (Blanton et al., submitted) Mapping nearby galaxies at Apache Point Observatory (MaNGA) survey, and build on the analysis presented in \cite{Belfiore2016a} (henceforth \textit{Paper I}) to define spatially resolved `quiescence' and make an unbiased census of star forming and quiescent regions within the z$\sim$0 galaxy population. This work is structured as follows. In Sec. \ref{sec2} we briefly recap the relevant properties of the MaNGA data and our spectral fits. In Sec. \ref{sec3} we summarise the main findings of \textit{Paper I}, including our new emission-line based galaxy classification, the importance of low ionisation emission-line regions (LIERs) in the context of understanding quiescence in galaxies and the stellar populations properties of the different galaxy classes. In Sec. \ref{sec4} we explore the role of LIER galaxies within the bimodal galaxy population, while in Sec. \ref{sec5} we gather evidence regarding the origin of the ionised gas in LIERs. Finally in Sec. \ref{sec6} we discuss possible evolutionary scenarios for LIER galaxies. We conclude in Sec. \ref{concl}.

Throughout this work redshifts, optical and UV photometry and stellar masses are taken from the MaNGA targeting catalogue, based on an extended version of the NASA Sloan Atlas (NASA-Sloan catalogue (NSA {\tt v1\_0\_1}\footnote{http://www.sdss.org/dr13/manga/manga-target-selection/nsa/}, \citealt{Blanton2011}). Stellar masses are derived using a \cite{Chabrier2003} initial mass function and using the {\tt kcorrect} software package (version {\tt v4\_2}, \citealt{Blanton2007}) with \cite{Bruzual2003} simple stellar population models. All quoted magnitudes are in the AB system and k-corrected to rest frame after correction for galactic extinction. Effective radii are measured from the SDSS photometry by performing a S\'ersic fit in the r-band. Single fibre spectroscopic data from the SDSS Legacy Survey data release 7, \citep{Abazajian2009} for the main galaxy sample targets \citep{Strauss2002} are referred to as legacy SDSS.
When quoting luminosities, masses and distances we make use of a $\Lambda$CDM cosmology with $\Omega_m=0.3$, $\Omega_\Lambda=0.7$ and $\rm H_0=70 \mathrm{ \ km^{-1} s^{-1}Mpc^{-1}}$.


\begin{figure*} 
\includegraphics[width=1.0\textwidth, trim=0 0 0 0, clip]{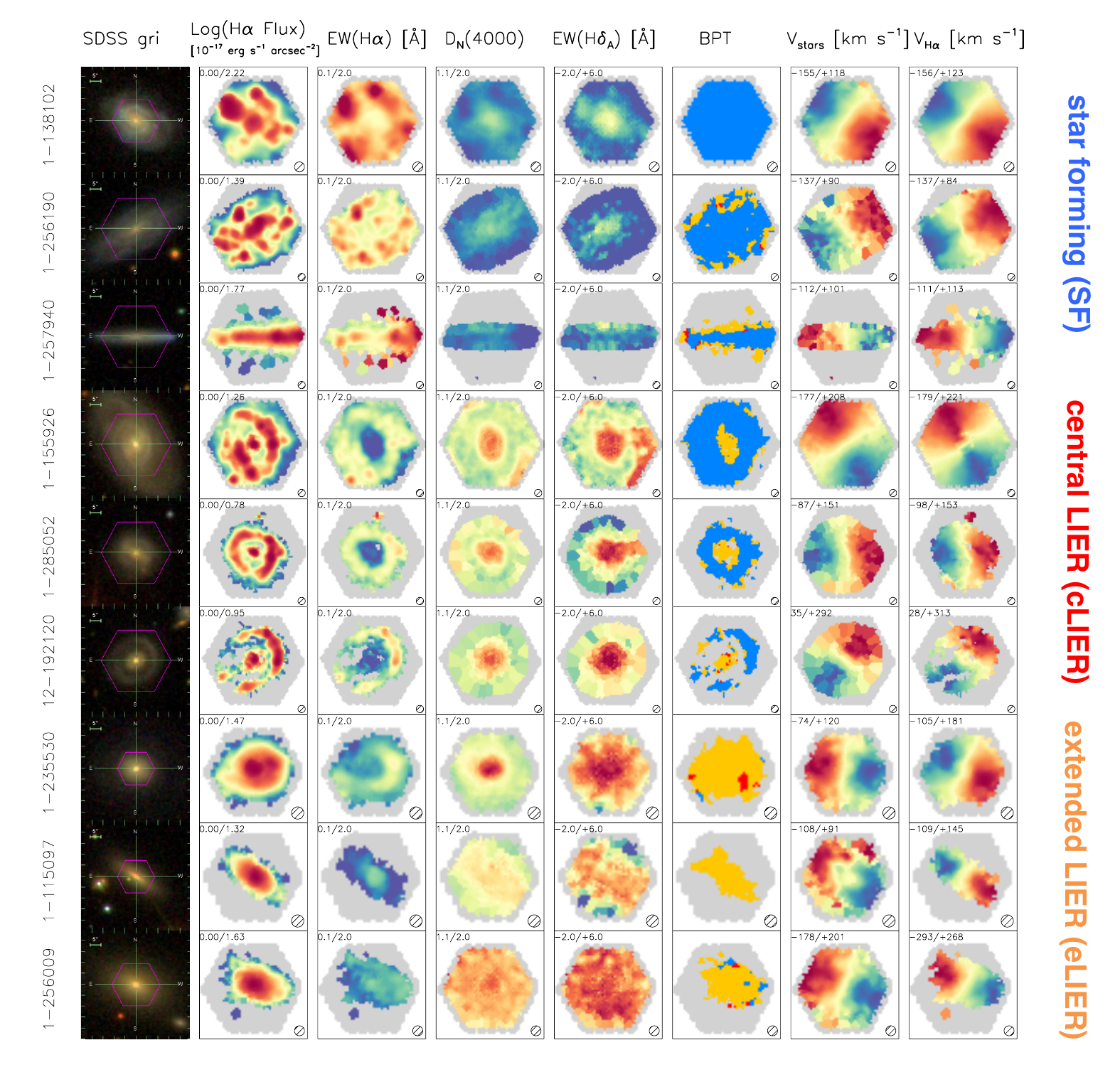}
\caption{Example of the different properties of SF (top three), cLIER (middle three) and eLIER (bottom three) galaxies. Each row represents a different galaxy (labelled with its MaNGA-ID on the left). The columns represent respectively (1) SDSS g-r-i composite image with MaNGA bundle superimposed, (2) MaNGA H$\alpha$ map, (3) MaNGA EW(H$\alpha$) map, (4) MaNGA map of $\rm D_N(4000)$, (5) MaNGA map of $\rm H\delta_A$, (6) BPT map of the MaNGA galaxies based on the [SII] BPT diagram. SF regions are colour-coded in blue, LIER in orange, Seyfert-like in red, (7) velocity field of the stars, (8) velocity field of the gas. For all maps, the colour bar increases from blue to red and the values corresponding to the minimum and maximum of the colour-bars are annotated on the top right corner of each map. The average PSF for the MaNGA data is shown in the bottom right corner of the maps.}
\label{fig2}
\end{figure*}

\section{The MaNGA data}
\label{sec2}

MaNGA \citep{Bundy2015, Yan2016a} is an IFS survey, making use of the SDSS 2.5m telescope at Apache Point Observatory \citep{Gunn2006}, aimed at obtaining spatially resolved spectroscopy for 10~000 nearby galaxies in the redshift range 0.01 $<$ z $<$ 0.15. Observations are obtained using purpose-built hexagonal integral field units (IFUs, \citealt{Drory2015}), assembled from 2$''$ diameter fibres, which are fed to the dual beam BOSS spectrographs covering the wavelength range 3600 and 10300 \AA\ with a spectral resolution R $\sim$ 2000 \citep{Smee2013}. The size distribution of the IFUs is optimised to match the size and density of galaxy targets on the sky (Wake at al., \textit{in prep.}), in order to cover 2/3 of the targets to 1.5 effective radii ($\rm R_e$) and 1/3 of the targets to 2.5 $\rm R_e$. The MaNGA sample is selected to be flat in i-band absolute magnitude. At each i-band magnitude (and thus approximately in each stellar mass bin) the MaNGA primary and secondary samples are selected to be volume limited within a prescribed redshift range.

Seventeen galaxy IFUs (ranging from 19-fibre IFU, 12$''$ on sky diameter, to 127-fibre IFU, 32$''$  on sky diameter) are observed simultaneously, together with a set of twelve 7-fibre minibundles used for flux calibration \citep{Yan2016} and 92 single fibres for sky subtraction. A three-point dithering pattern is used during observations to compensate for light loss and obtain a uniform point spread function (PSF, \citealt{Law2015}). The MaNGA data were reduced using version {\tt v1\_3\_3} of the MaNGA reduction pipeline \citep{Law2016a}. The wavelength calibrated, sky subtracted and flux calibrated MaNGA fibre spectra (error vectors and mask vectors) and their respective astrometric solutions are combined to produce final datacubes with pixel size set to 0.5$''$. The median PSF of the MaNGA datacubes is estimated to have a full width at half-maximum (FHWM) of 2.5$''$. The MaNGA sample has been demonstrated to be well-suited for studies of gas physics and stellar population properties of nearby galaxies as shown in preliminary analysis of data from the MaNGA prototype instrument \citep{Belfiore2015, Li2015, Wilkinson2015}.

Physical parameters of the continuum and the emission lines are obtained via a customised spectral fitting procedure described in \textit{Paper I} (Sec. 2.2). In short, spaxels are binned for optimising the S/N ratio in the continuum, which is then fitted with a set of MIUSCAT simple stellar population templates \citep{Vazdekis2012} via the penalised pixel fitting routine \citep{Cappellari2004}. Subsequently a new binning is performed, optimising the H$\alpha$ S/N, which is used to fit nebular emission lines. Emission lines are fitted as sets of Gaussians (one per line) with a common velocity. Further constrains are imposed on lines originating from the same ion.


\section{LIER emission in galaxies: classification and open questions}
\label{sec3}

\subsection{Paper I: A revised classification scheme based on spatially resolved diagnostic diagrams}
\label{sec3.1}

In \textit{Paper I} we investigated the spatial distribution and properties
of galactic regions that are classified as `LINERs' (low ionisation nuclear emission-line regions) in the conventional [SII]$\lambda\lambda 6717, 6731$/H$\alpha$ versus [OIII]$\lambda 5007$/H$\beta$ BPT \citep{Baldwin1981} diagnostic diagram.
We showed that such regions are not nuclear, but distributed on large (kpc) scales, hence motivating the relabelling of these regions as `LIERs' (i.e. dropping the `N', which stood for `nuclear', from the historical acronym).
We made use of different lines of evidence to argue against the AGN origin of LIER emission,
instead supporting the scenario where LIERs are caused by photoionisation from a diffuse
ionising background produced by hot evolved stars. 

More specifically we summarise the main conclusions of  \textit{Paper I} as follows:

\begin{enumerate}
\item{LIER emission is spatially strongly correlated with absence of young stellar populations (as traced by stellar indices like $\rm D_N(4000)$ and $\rm H\delta_A$).}
\item{Post asymptotic giant branch (pAGB) stars have been shown to produce the required hard ionising spectrum necessary to excite LIER emission \citep{Binette1994, Stasinska2008}. Stellar evolution models including the pAGB phase demonstrate that the pAGB hypothesis is energetically viable for virtually all LIERs, as the observed EW(H$\alpha$) lie in the same range as model predictions (0.5 - 3.0 \AA).}
\item{Although weak AGN may appear as true LINERs when observed on scales of 100 pc or less, a typical low luminosity AGN does not emit enough energy to be the dominant contributor to the ionising flux on the kpc scales probed by the 3$''$ SDSS fibre (from the legacy SDSS, $\rm <z> \sim$ 0.1) or by MaNGA ($\rm <z> \sim$ 0.03)}
\item{In LIER galaxies not contaminated by SF, line ratios sensitive to the ionisation parameter ([OIII]$\lambda$5007/[OII]$\lambda$3727 and partially [OIII]$\lambda$5007/H$\beta$) show flat or very shallow profiles over radial scales of tens of kpc, at odds with the steeper profiles predicted in the AGN scenario.}
\item{The distribution of EW(H$\alpha$) in LIERs is flat as a function of galactocentric radius and presents a
remarkably small scatter, as expected
in the case of nebular lines being excited by hot evolved stars associated with the old
stellar population. On the other hand, in the AGN scenario, a large scatter is expected in the EW(H$\alpha$) distribution, as line and stellar continuum emission would originate from independent sources.
}
\end{enumerate}

While previous work has associated LIER emission in red (elliptical) galaxies with pAGB stars \citep{Binette1994, Sarzi2010, Yan2012, Singh2013}, we note that
\textit{Hubble Space Telescope} imaging of very nearby elliptical galaxies struggles to account for the number of UV bright (especially pAGB) stars predicted by stellar population models \citep{Brown1998, Brown2008, Rosenfield2012}. If models have substantially overestimated the occurrence of pAGB stars, other hot evolved stars (most notably extreme horizontal branch stars) may represent a more sizeable contribution to the ionising photon budget leading to LIER emission. For the purposes of this paper, the exact nature of the evolved stellar sources powering the LIER emission does not affect any of the conclusions.

In this framework LIERs are direct tracers of regions dominated by old stellar populations where, despite the presence of gas, star formation is no longer taking place. Importantly, LIERs appear both in red sequence galaxies, which host no current star formation, but also in the central regions of spiral galaxies, where star formation actively takes place at larger galactocentric radii (as recently highlighted by the contributions from the CALIFA survey, \citealt{Singh2013, Gomes2016}). We refer to these two classes of galaxies `extended LIER' (eLIER) and `central LIER' (cLIER) respectively, with the absence or presence of SF regions being the discriminating factor. Examples of galaxies in each class are given in Fig. \ref{fig2}, together with maps of their line emission (H$\alpha$ flux and EW), age-sensitive stellar population indices ($\rm D_N(4000)$ and $\rm EW(H\delta_A$)) and BPT classification (blue corresponding to SF, orange to LIER and red to Sy-like regions).

In this work we make use of 583 galaxies selected from the sample described in \textit{Paper I}. In particular, we exclude Seyfert galaxies,
mergers and galaxies with an ambiguous classification and classify all the remaining galaxies as described in \textit{Paper I}, i.e.:

\begin{enumerate}
\item{\textit{Line-less galaxies}: No detected line emission (average EW(H$\alpha$) $< $1.0 \AA\ within 1 $\rm R_e$).}
\item{\textit{Extended LIER galaxies (eLIER)}: galaxies dominated by LIER emission at all radii where emission lines are detected. No evidence for any star forming regions.}
\item{\textit{Central LIER galaxies (cLIER)}: galaxies where LIER emission is resolved and located in the central
regions, while ionisation from star formation dominates at larger galactocentric distances.}
\item{\textit{Star forming galaxies (SF)}: galaxies dominated by star formation in the central regions and at
all radii within the galaxy disc.}
\end{enumerate}
As discussed in \textit{Paper I}, other `ionisation morphologies' are extremely rare. For example, we find no galaxy with a line-less bulge and a star forming disc. The 583 galaxies used in this paper subdivide in the classes above as follows: 314 SF, 57 cLIER, 61 eLIER and 151 line-less.

\begin{figure*} 
\includegraphics[width=\textwidth,  trim=0 200 0 100, clip]{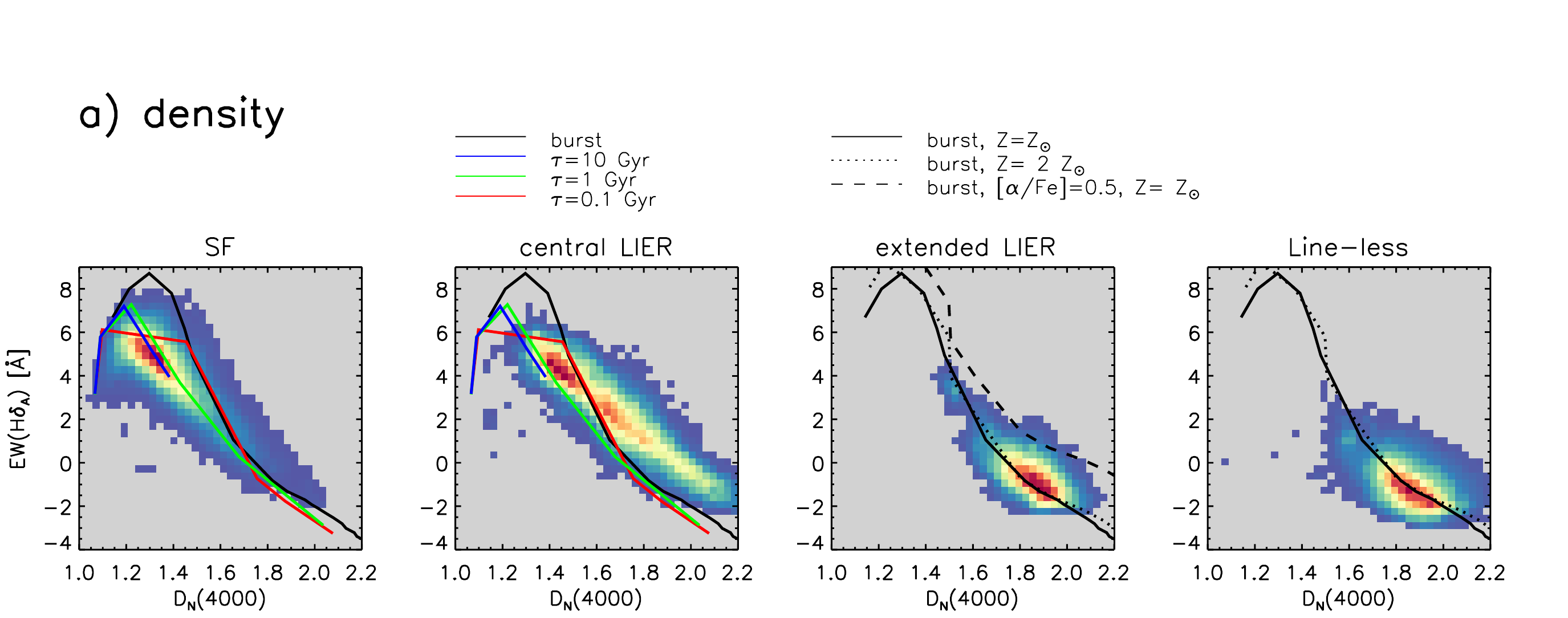}
\includegraphics[width=\textwidth,  trim=0 200 0 120, clip]{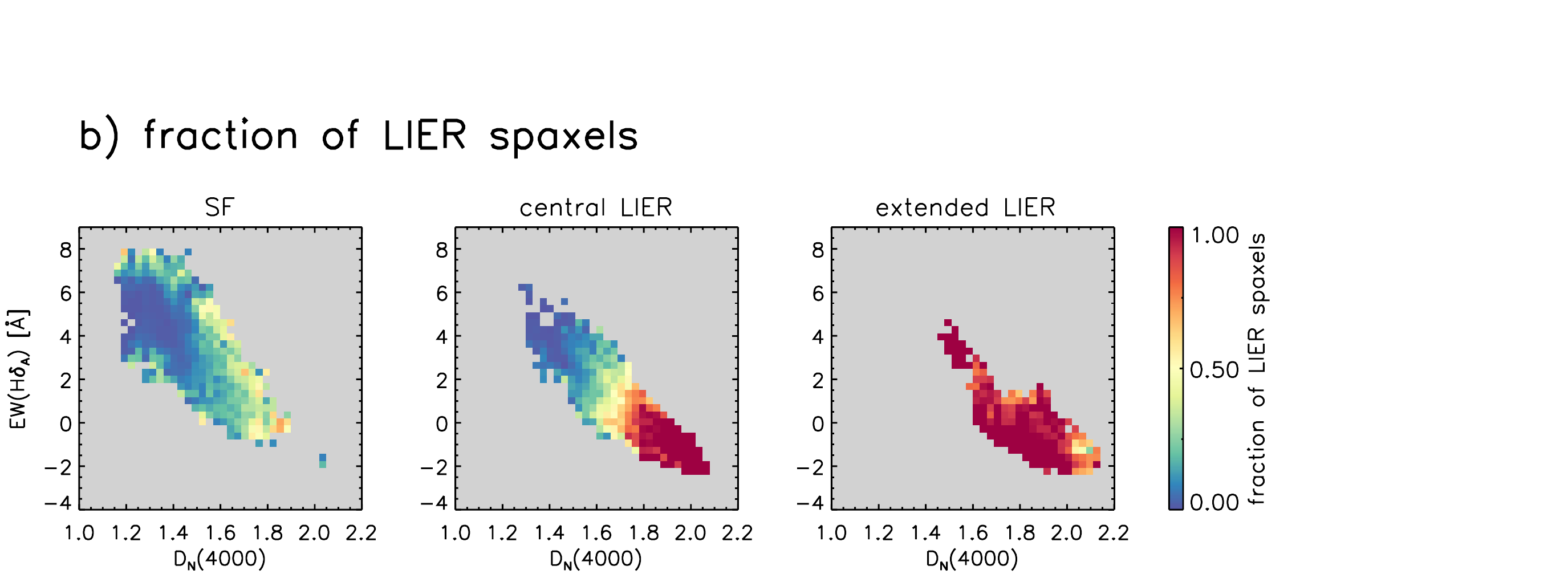}
\caption{a) Density plot in the $\rm D_N(4000)$ versus $\rm EW(H\delta_A)$ plane for spaxels in the different galaxy classes (SF, cLIER, eLIER and line-less) defined in this work. Different burst and continuous SFR models based on \protect\cite{Maraston2011} stellar population models are overplotted (details in the legend).
b) Same as a) but with the colour-coding reflecting the fraction of spaxels classified as LIER in each region of the $\rm D_N(4000)$ versus $\rm EW(H\delta_A)$ plane.}
\label{fig2.1}
\end{figure*}

\subsection{From nebular lines to stellar indices: tracing star formation and quiescence}
\label{sec3.3}

The realisation that LIER emission is not related to nuclear activity, but simply a
consequence of presence of a diffuse UV background due to evolved stellar populations, leads us
to reconsider the conventional definition of quiescence as absence of nebular line emission.
Namely, both LIER and line-less regions (EW(H$\alpha) <$ 1 \AA) can ben defined as `quiescent' (or `passive'), to highlight the fact that they do not host current star formation as traced by the dominant component of line emission.

An alternative definition of quiescence may rely on probes of star formation on longer
time-scales, such as the so-called 4000-\AA\ break ($\rm D_N(4000)$, \citealt{Balogh1999}) that
is a sensitive tracer of the age of the stellar population. More specifically,
while $\rm D_N(4000)$ is also sensitive to metallicity effects (for stellar population older than 1 Gyr), the combination of $\rm D_N(4000)$ and the stellar index $\rm EW(H\delta_A)$ is less sensitive to the effect of metallicity and has been used in the literature as a probe for the galaxy's star formation history \citep[e.g.][]{Kauffmann2003b, Kauffmann2003c, Li2015}. In \textit{Paper I} we have demonstrated that a selection on $\rm D_N(4000)$ or EW(H$\alpha$) is generally equivalent to our emission-line base classification, as LIERs are characterised by the absence of young stars.

In Fig. \ref{fig2.1}(a) we plot the MaNGA spaxels divided by galaxy class (SF, cLIER, eLIER and line-less) in the $\rm D_N(4000)$ versus $\rm EW(H\delta_A)$ plane. Model tracks, based on \cite{Maraston2011} stellar population models, are shown for a burst of star formation (with solar metallicity, solid black line in all panels) and three models with continuous, exponentially declining star formation (where the SFR decline time-scale, $\tau$, is varied in the range 0.1-10 Gyr). Except for the increase in $\rm EW(H\delta_A)$ observed in the burst model at young ages ($<$ 1 Gyr), stellar population age increases going from top-left to bottom-right. The black dotted models in the eLIER and line-less panels correspond to a burst of twice solar metallicity, and confirm that metallicity does not substantially affect the position in the $\rm D_N(4000)$ versus $\rm EW(H\delta_A)$. In all panels, the burst model defines an envelope in $\rm D_N(4000)$-$\rm EW(H\delta_A)$, with continuous SFR models occupying regions at lower $\rm D_N(4000)$ for a fixed $\rm EW(H\delta_A)$ value.

Continuous star formation models are an excellent representation of the data for SF galaxies, with the model with $\tau$=10 Gyr passing through the locus of the data. The data for cLIERs is bimodal in the $\rm D_N(4000)$ versus $\rm EW(H\delta_A)$ plane, spanning the full range of stellar ages. The peak at $\rm D_N(4000) \sim 1.4$ is well-fitted by continuous SFR models with a decline time-scale of several Gyr. However, the second peak  ($\rm D_N(4000) \sim 1.7$) and the high-$\rm D_N(4000)$ tail of the distribution is harder to fit with the available models because it lies above the burst model track. Dust can have an effect on the computed indices, especially $\rm D_N(4000)$, and contribute to resolving the observed discrepancy. 

The stellar populations of eLIERs and line-less galaxies are indistinguishable in the $\rm D_N(4000)$-$\rm EW(H\delta_A)$ plane and are well-fitted by the burst model at old ($>$ 10 Gyr) ages. While the locus of the data does not require significantly $\alpha$-enhanced stellar populations, $\alpha$-enhancement can successfully explain the observed spread to higher $\rm EW(H\delta_A)$ at fixed $\rm D_N(4000)$ (dashed model track in eLIER panel, \citealt{Thomas2004, Thomas2011}).

In Fig. \ref{fig2.1}(b), we show the same plot, but colour-coded by the fraction of spaxels classified as LIER across the $\rm D_N(4000)$ versus $\rm EW(H\delta_A)$ plane. The results demonstrate the strong correspondence between LIERs and old stellar populations, with cLIERs showing a transition from LIER-dominated central regions to SF-dominated outskirts. 

The tight relation between line emission and stellar populations leads us to further investigate the emission-line based galaxy classification in the context of the broader physical properties of galaxies. In this work we wish to address the question of how emission-line-based galaxy categories (SF, cLIER, cLIER and line-less) fit within the overall galaxy bimodality in terms of their fundamental properties (colours/morphology/kinematics). For example, can we interpret the SF, cLIER, eLIER and line-less galaxies sequence of increasing `quiescence' as an evolutionary sequence?
Where does the gas in eLIER galaxies come from and what makes these galaxies different from line-less galaxies?
Here, we attempt to answer these questions by taking advantage of both the MaNGA data and legacy SDSS photometry and spectroscopy.

%
%
%
\section{LIER galaxies within the bimodal galaxy population}
\label{sec4}

\begin{figure*} 
\includegraphics[width=0.8\textwidth, trim=50 40 50 80, clip]{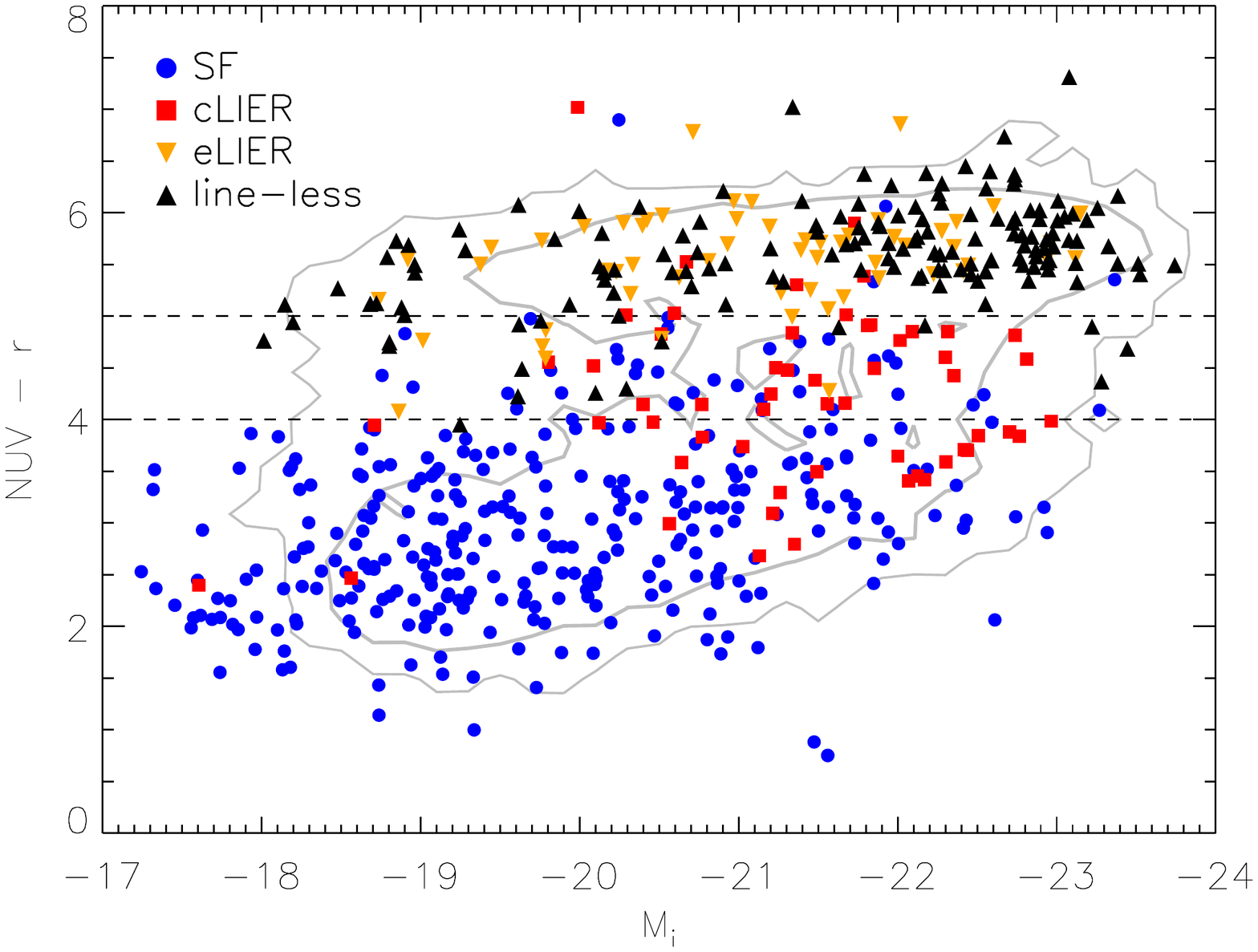}
\caption{MaNGA galaxies in the $NUV - r$ versus $M_i$ (colour-magnitude) diagram. Galaxies are divided according the emission line-based classification introduced in \textit{Paper I} (see also Section \ref{sec2} of this work). Contours represent galaxy density in the MaNGA parent sample, while dashed lines provide a simple division between the red sequence ($NUV - r \rm> 5$), green valley ($4<$ $NUV - r \rm< 5$) and blue cloud ($NUV - r \rm< 4$). The green valley is preferentially occupied by cLIER galaxies, while eLIER and line-less galaxies are mostly found on the red sequence ($NUV - r \rm> 5$).}
\label{fig4.1}
\end{figure*}

\subsection{Integrated colours and star formation rates}
\label{sec4.1}

The physical origin of the colour bimodality in the galaxy population lies in the idea of using galaxy colour as `clocks'. Both UV and blue
colours evolve quickly once O and B stars disappear from the galaxy spectrum. In particular, NUV emission declines more slowly than H$\alpha$ ($\rm
\sim 10^6-10^7 ~yr$), but faster than blue optical continuum ($\rm \sim 10^9 ~yr$), making UV-optical colours a sensitive probe to the fading of young stellar populations.

Fig. \ref{fig4.1} shows the $NUV - r$ versus M$_i$ (i-band absolute magnitude) diagram for the current MaNGA sample, with galaxies sub-divided according to the emission-line-based classification scheme introduced in this work. We observe that SF galaxies constitute the vast majority of the blue cloud ($NUV - r < 4$), while line-less and eLIER galaxies are concentrated on the red sequence ($NUV -r > 5$). The most striking feature of Fig. \ref{fig4.1} is that a large fraction of cLIER galaxies ($\sim$40 \%) are found in the green valley ($4<NUV - r<5$), with the remaining ones also lying at intermediate UV-optical colours (the mean $NUV - r$ for cLIERs is 4. 1 with a standard deviation of 1.0 mag). \textit{The prevalence of cLIER galaxies in the green valley is not surprising, since the integrated colours are affected by both the light from the red, old, LIER-like central regions and light from the blue, star forming disc, thus generating the observed intermediate integrated colours.} The separation of cLIER galaxies from the blue cloud is less evident when using optical colours (for example $u -r$ or $g - r$ instead of $NUV - r$). 
Indeed, using $g-i$ colours, cLIER galaxies are consistent with lying on the red sequence, and have equivalent colour to eLIERs and line-less galaxies. Further details on the UV-optical colours of different galaxy classes are summarised in Table \ref{table_data}.

\begin{figure*} 
\includegraphics[width=0.8\textwidth, trim=40 30 50 50, clip]{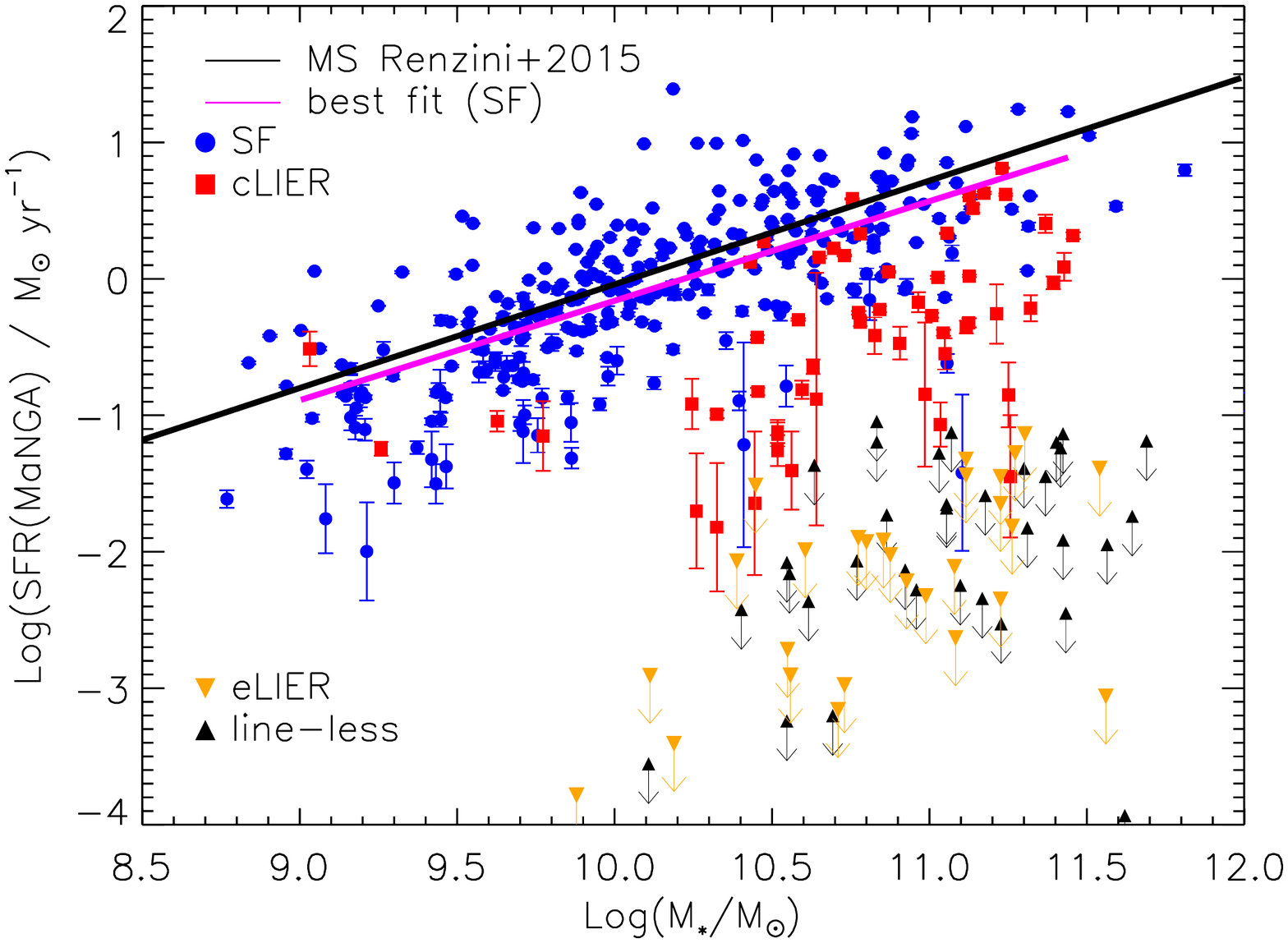}
\caption{The SFR-M$_\star$ plane for MaNGA galaxies. The SFR from the extinction-corrected H$\alpha$ flux by summing the flux in spaxels classified as SF. The black line represent the `main sequence' for SF galaxies from legacy SDSS spectroscopy \protect\citep{renzini2015}. The magenta line represents the best fitting `main sequence' using the SF galaxies in the MaNGA sample.}
\label{fig4.12}
\end{figure*}

Fig. \ref{fig4.12} presents an even more striking illustration of the transition nature of cLIER galaxies, by representing the MaNGA galaxies in the SFR-M$_\star$ plane. The SFR is calculated from the extinction-corrected H$\alpha$ flux within the MaNGA bundle and using the \cite{Kennicutt1998} conversion factor. The procedure used to correct the SFR for LIER emission is based on the assumption that the total H$\alpha$ flux can be decomposed into a contribution from star formation (with line ratios typical of star forming regions) and from LIERs/diffuse ionised gas (with line ratios typical of LIERs). The details of the correction procedure are presented in Appendix \ref{AppA}. This procedure generates SFR estimates that are on average indistinguishable from the SFR computed using only the spaxels classified as star forming using the [SII] BPT diagram. For LIER regions, only a small fraction of the H$\alpha$ flux can be attributed to underlying star formation, and we thus consider the SFR for eLIER and line-less galaxies as upper limits. We do not apply any aperture correction to the MaNGA-derived SFR. Errors on the SFR are estimated by propagating the uncertainties in the flux and extinction correction. 

Fig. \ref{fig4.12} demonstrates that SF galaxies lie on the `main sequence' \citep{Brinchmann2004, Salim2007,Peng2010, renzini2015} while the cLIER galaxies lie systematically below it, although above the passive galaxy population, consisting of eLIERs and line-less galaxies.
For reference we plot the `main sequence' at $\rm z \sim 0$ from \cite{renzini2015} (black solid line in Fig. \ref{fig4.12}), obtained from legacy SDSS spectroscopy, and the best-fitting line to the SF galaxy population using the current MaNGA sample (with $\rm 9 < log(M_\star/M_\odot) < 11.5$, magenta line in Fig. \ref{fig4.12}), given by 
\begin{equation}
\rm log\left( \frac{SFR}{M_\odot yr^{-1}} \right)= -7.5 ~\pm 0.2 + (0.73 ~\pm 0.02) ~log\left( \frac{M_\star}{M_\odot} \right)
\end{equation}
Considering the differences in sample selection and methodology for calculating stellar mass and
SFR, results from legacy SDSS are in excellent agreement with the MaNGA result presented here.
At a fixed stellar mass, cLIER galaxies lie on average 0.8 dex below the main sequence defined
by the SF population, although they actually spread across the
entire green valley with large scatter.

Summarising, the key results from this section are the following: (1) eLIER and quiescent line-less galaxies are indistinguishable in terms of colours; (2) most green valley galaxies are not such because of dust reddening or globally intermediate stellar populations, but are associated with galaxies (cLIERs) in which quiescent (non-SF) central regions co-exist with outer star formation. These galaxies lie in the transition region between the star formation main sequence and quiescent galaxies. 

\begin{table*}
\caption{Classification statistics of emission line galaxies in the $NUV - r$ colour-magnitude diagram and mean ($\pm$ standard deviation) for selected UV and optical colours. Appropriate volume weights have been applied to correct for the MaNGA selection function.}
\centering
\begin{tabular}{ c c c c c c c}

\hline 
Galaxy class & Blue cloud & Green Valley & Red sequence  				& $NUV -r$  &$ u -r $& $g -r$ \\ 
    		     & $NUV - r \rm< 4$  & $4<$ $NUV - r \rm< 5$ & $NUV - r \rm> 5$ & $\mu \pm \sigma$ &$\mu \pm \sigma$  &$\mu \pm \sigma$  \\
\hline 
%
SF		&  92\%	& 7\%   &  1\% 		&  2.8 $\pm$ 0.8	&  	1.6 $\pm$ 0.4	&	0.55 $\pm$ 0.16	\\
cLIER	& 48\%	& 39\%   & 13\% 	&  4.1 $\pm$ 1.0	&	2.2$\pm$ 0.3	&	0.74 $\pm$ 0.10	\\ 
eLIER	& 0  		& 19\%  & 81\% 	&  5.4 $\pm$ 0.6	&	2.5 $\pm$ 0.2	&	0.78 $\pm$ 0.09	\\ 
line-less	& 1\%  	& 14\%   & 85\% 	&  5.5 $\pm$ 0.7	&	2.4 $\pm$ 0.2	&	0.76 $\pm$ 0.09	\\ 
\hline

\end{tabular}
\label{table_data}
\end{table*}
%
%
%
\begin{figure*} 
\includegraphics[width=0.49\textwidth, trim=50 40 50 50, clip]{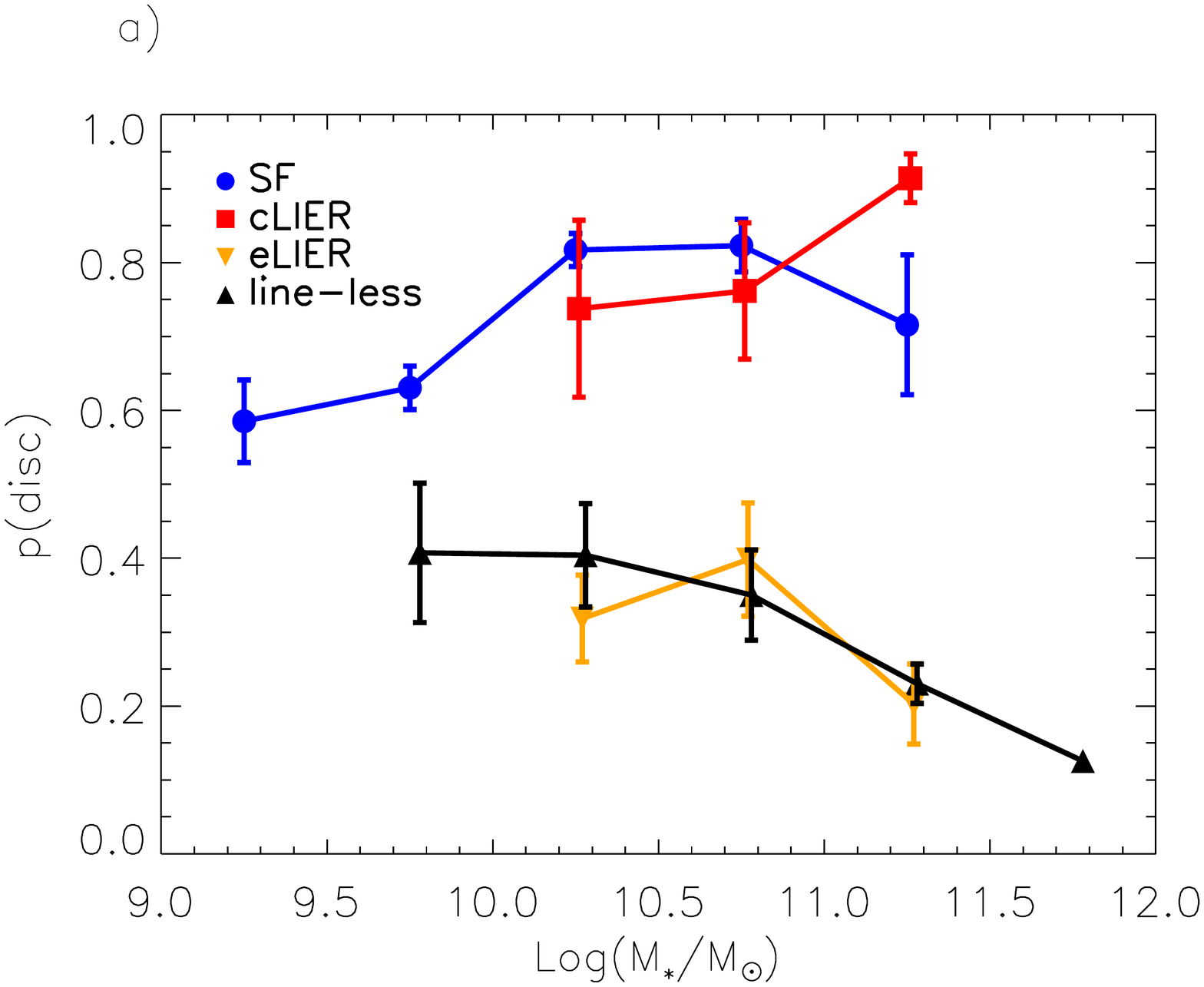}
\includegraphics[width=0.49\textwidth, trim=50 40 50 50, clip]{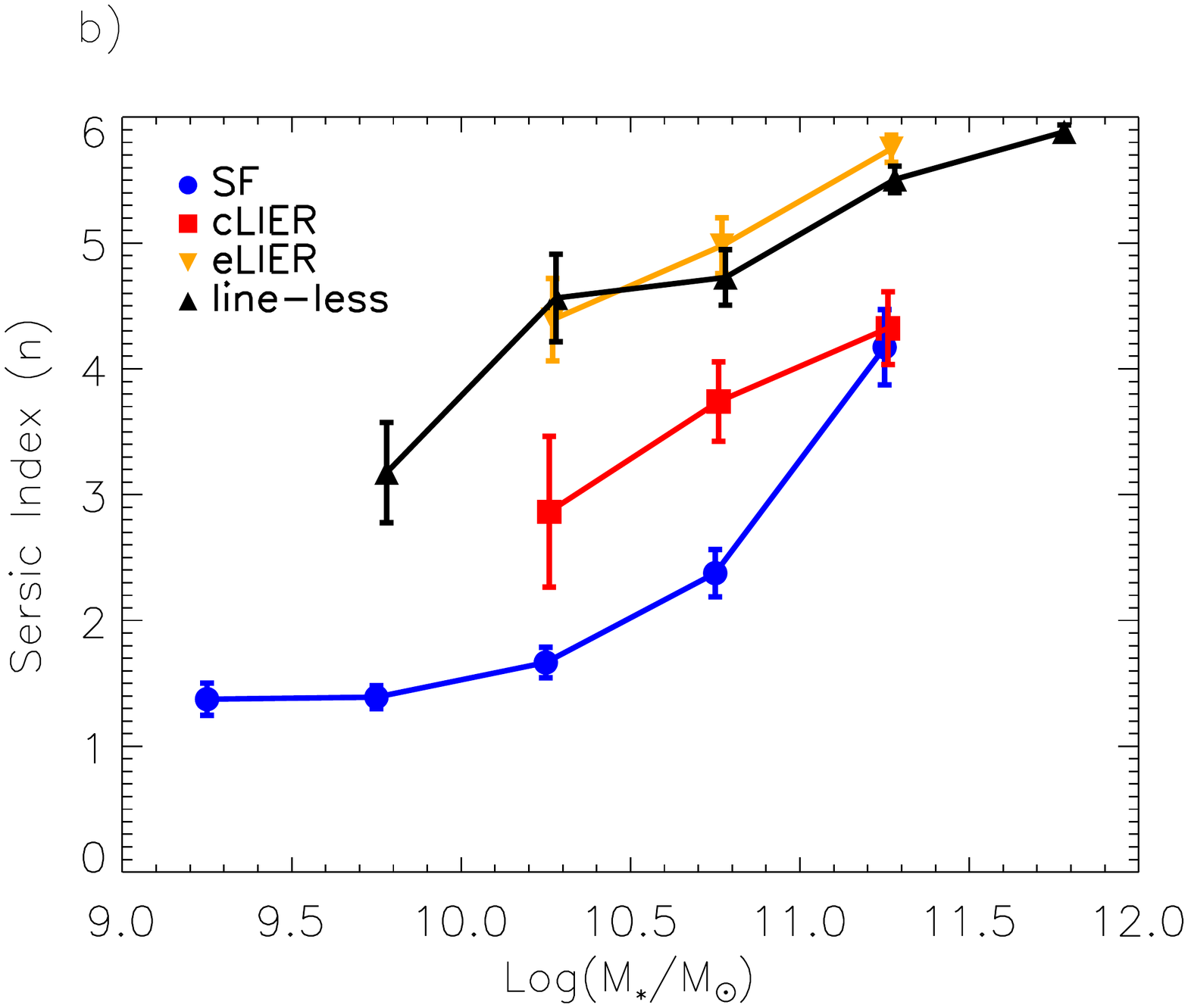}
\includegraphics[width=0.49\textwidth, trim=50 40 50 50, clip]{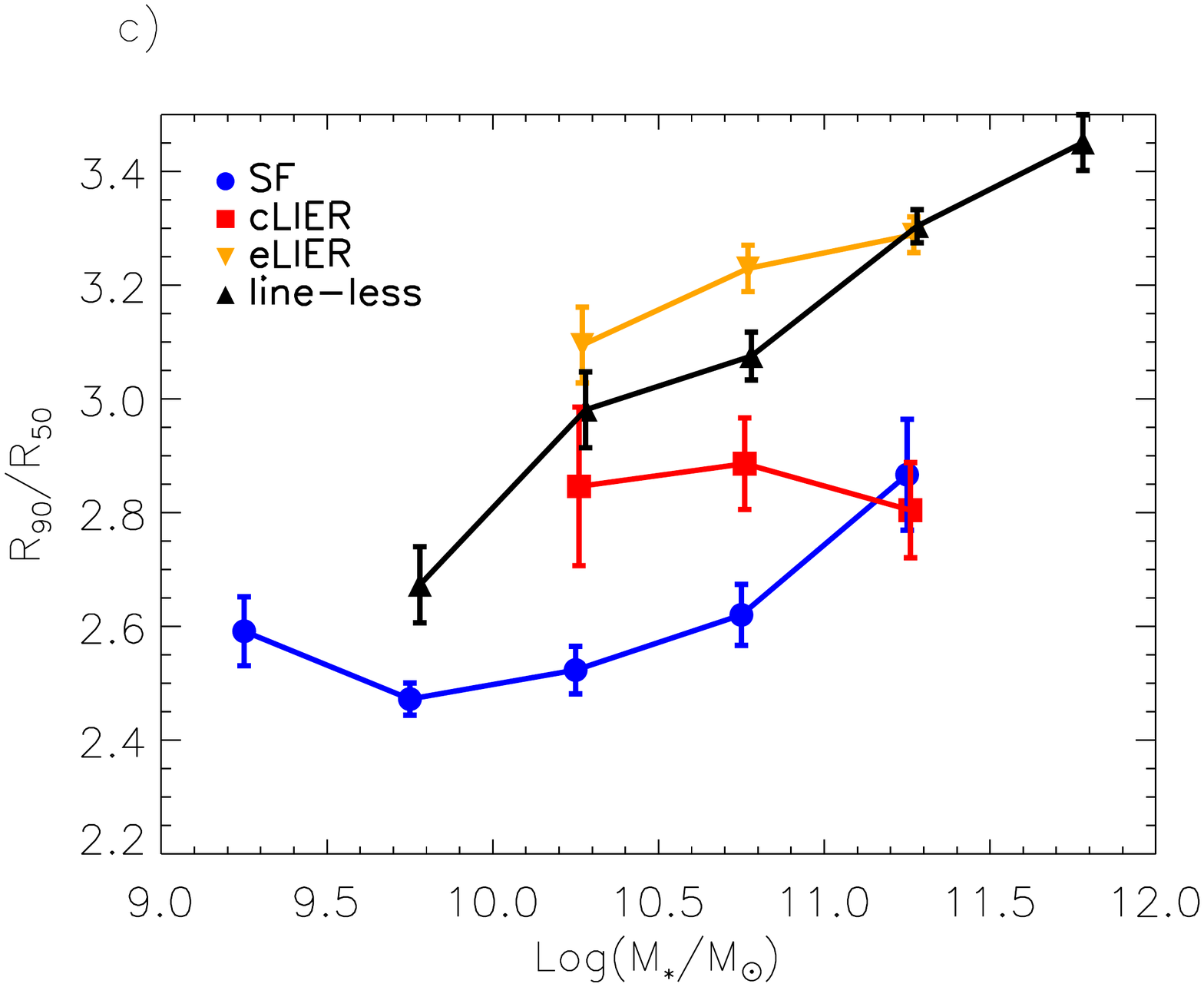}
\includegraphics[width=0.49\textwidth, trim=45 40 50 50, clip]{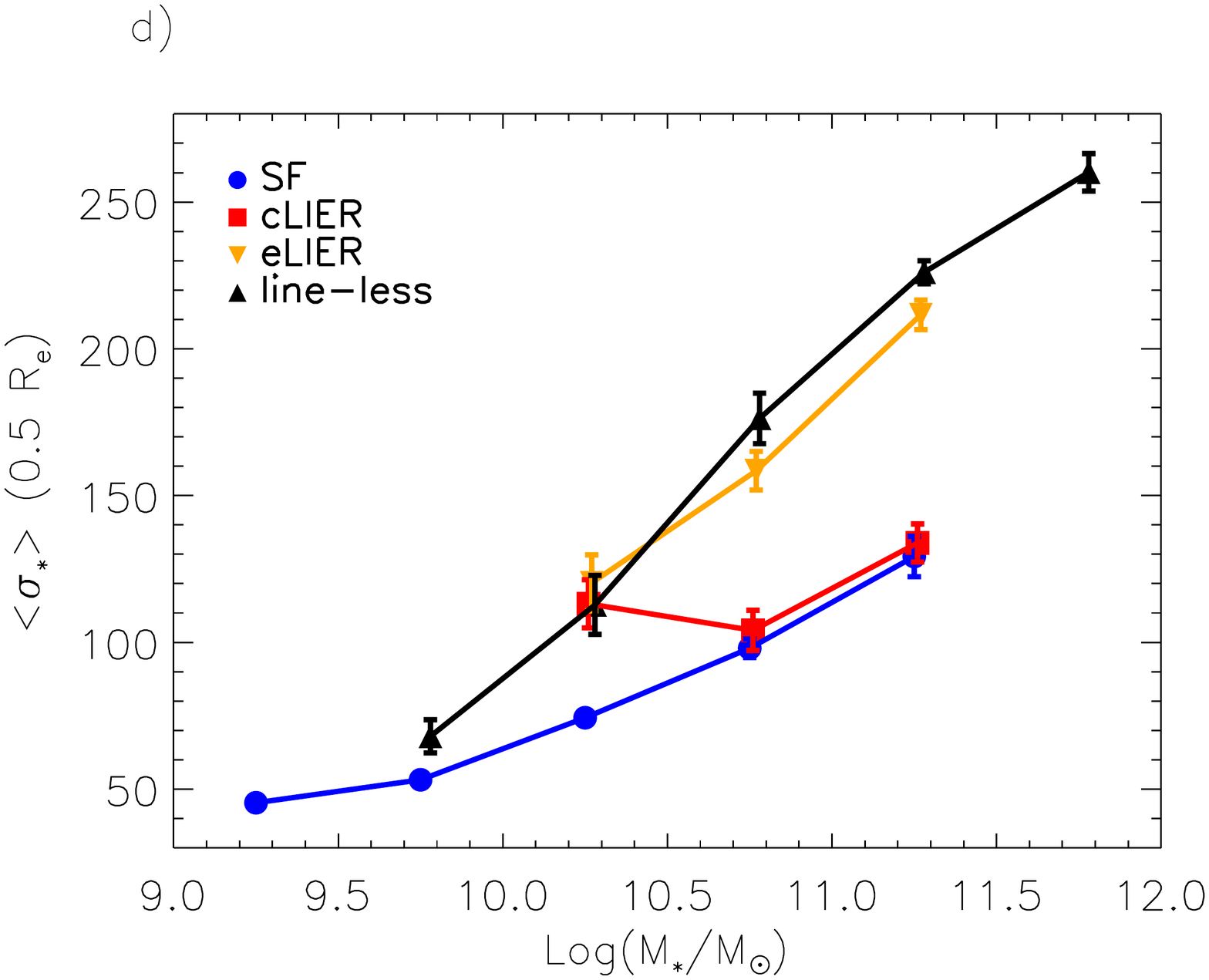}

\caption{\textit{a)} Average disc vote fraction from Galaxy Zoo 2 \protect\citep{Willett2013} in bins of stellar mass for MaNGA galaxies of different classes. The vote fraction can be interpreted as a probability and has been statistically de-biased to account for the increasing difficulty of detecting fainter morphological features at higher redshift following \protect\cite{Willett2013}. \textit{b), c)} Average S\'ersic index and concentration (defined as $\rm R_{90}/R_{50}$, where R are Petrosian radii) in bins of stellar mass. S\'ersic parameters and Petrosian radii are obtained from the NSA \protect\citep{Blanton2011} and are based on legacy SDSS r-band imaging. \textit{d)} Average stellar velocity dispersion within 0.5 $\rm R_e$ in bins of stellar mass. Velocity dispersions are calculated from the MaNGA data and $\rm R_e$ is the S\'ersic r-band effective radius from the NSA catalogue.}
\label{fig4.3}
\end{figure*}

%
\subsection{Morphological type}
\label{sec4.2}

The traditional morphological dichotomy between early type (ETG, E and S0) and late type galaxies (LTG, spirals), as implemented in the revised Hubble classification \citep{Sandage1961}, is based on the presence or absence respectively of spirals arms and/or dust lanes when galaxies are seen edge on. 
Subsequent revisions to this classification, including recent advanced made possible by IFS observations, have conclusively demonstrated that a large fraction of ETG contain discs seen at a variety of inclinations (dubbed as `fast rotators' by the ATLAS$^{\rm 3D}$ project). Only the most massive end of the ETG population is populated by true dispersion dominated systems (`slow rotators'), with very little ordered angular momentum \citep{Emsellem2007, Emsellem2011, Cappellari2011}. Extensions to higher redshift \citep{Bundy2010} seem to confirm this general picture.

Visual morphological classifications from the Galaxy Zoo 2 project \citep{Lintott2008, Willett2013}, based on legacy SDSS imaging, were used to subdivide MaNGA galaxies into ETG and LTG. In particular, we make use of the detailed classification from Galaxy Zoo 2, which asked its volunteer citizen scientists to identify several morphological features (spirals, bars etc) in SDSS three-colour image cutouts. Individual classifications are processed as detailed in \cite{Willett2013} and statistically `de-biased' to take into account the fact that is it harder to identify fine morphological features for smaller, fainter galaxies. The nature of the Galaxy Zoo vote fractions means that they can be interpreted as conditional probabilities, although one needs to take into account that not all parameters are estimated for all galaxies, given the decision tree nature of the Galaxy Zoo 2 classifications. Vote fractions for certain questions (hence features) are only computed if previous questions in the decision tree are `well-answered' (as detailed in \citealt{Willett2013}, section 3.3 and table 3).

In Fig. \ref{fig4.3} we plot the median trends in bins of stellar mass for the de-biased probability of the presence of disc or other feature\footnote{Task 1 in Galaxy Zoo 2, addressing the question: \textit{Is the galaxy simply smooth and rounded, with no sign of a disc?}} in our MaNGA galaxy sample, classified according to the emission-line based classification. In order to study trends as a function of stellar mass we correct for the effect of the MaNGA selection function. The error bars represent the error on the mean in each mass bin. The results from the Galaxy Zoo 2 classifications demonstrate that eLIER and line-less galaxies have mutually consistent morphologies and are visually classified as ETGs. SF and cLIER galaxies, on the other hand, also have mutually consistent disc fractions and are consistent with being LTGs. Although outliers are present in either direction (SF galaxies with no disc features or line-less galaxies classified as discs), these do not dominate the sample averaged trends.
 
Interestingly, previous work based on Galaxy Zoo has highlighted the fact that when galaxies are classified into a red sequence and blue cloud using optical colours, the red sequence host a substantial fraction of red (`anemic') spirals \citep{Bamford2009, Masters2010}. cLIER galaxies, because of their substantially redder optical colours than the mean of the blue cloud (see Table \ref{table_data}) and their disc-like morphology, may appear as `red spirals' in integrated colours. This class of galaxies represents a natural link between normal spirals and lenticular (S0) galaxies, as suggested in the revised `trident' morphological classification \cite{VandenBergh1976} and further updated by \cite{Cappellari2011}.

Further support towards a relation between the anemic spiral class and cLIERs is given by the fact that anemic spirals in single-fibre spectroscopy \citep{Masters2010} are often classified as LIERs, have lower EW(H$\alpha$) than SF galaxies, present older mean stellar ages \citep{Tojeiro2013} and are found to be an insignificant fraction of the total spiral population for $\rm M_\star < 10^{10}~M_\odot$ \citep{Wolf2009} while representing as much as 40\% of the spiral population at the highest masses ($\rm M_\star>10^{10.8}~M_\odot$, \citealt{Masters2010}).

%
\subsection{The importance of the bulge}
\label{sec4.3}

Morphological features such as spiral arms and bars only capture one aspect of the complex problem of morphological classification of galaxies. 
Another characteristic generally associated with the study of galaxy morphology is the presence and importance of a central spheroidal component (the bulge). Unfortunately, although parametric bulge-disc decomposition is a well-established technique, studies based on photometry alone are limited by the well-known intrinsic degeneracies in the method \citep{Rybicki1987, Gerhard1996}. Although potentially even more biased, `cheaper' proxies such as the `concentration' index (here defined as $\rm C = R_{90}/R_{50}$ where R are Petrosian radii) or a galaxy-averaged S\'ersic index have often been used on large photometric samples to provide an observationally more well-defined estimate of the importance of the bulge/spheroidal component. As a complement to these photometric indicators we also study the central stellar velocity dispersion extracted from the MaNGA data, which is a direct proxy for the depth of the bulge gravitational potential.

In this context, it is important to note that, if the bulge is defined kinematically, the classical lenticular (S0) morphological class is found to span a wide range in bulge fraction, forming a sequence essentially parallel to the classical increase in bulge fraction from Sc to Sa for LTGs \citep{Cappellari2011b, Krajnovic2013, Cappellari2016}. Indeed, it is arguable whether the traditional S0 morphological classification is in itself meaningful, as IFS surveys have demonstrated that the photometry alone is a very poor predictor of the degree of rotational support for ETG, with S0 likely to be classified as classical dispersion dominated ellipticals when viewed at low inclinations \citep[e.g][]{Emsellem2011}. 

In Fig, \ref{fig4.3}b), c), d) we show the average S\'ersic index, concentration and stellar velocity dispersion within 0.5 $\rm R_e$ ($<\sigma_\star>$) as a function of galaxy class in bins of stellar mass. In each bin, an average value is plotted only if there are more than four galaxies in the bin. As expected from their early type morphology, both line-less and eLIER galaxies show higher S\'ersic indices, concentrations and $<\sigma_\star>$ than cLIER and SF galaxies in all stellar mass bins. cLIERs have S\'ersic indices, concentrations and $<\sigma_\star>$ intermediate between eLIER/line-less and SF galaxies, consistent with the idea of being the subset of disc-dominated galaxies with largest bulge component at each stellar mass. For purely SF galaxies S\'ersic index, concentration and $<\sigma_\star>$ all increase as a function of mass, matching those of cLIER galaxies in the mass bin $\rm 11.0 < log(M_\star/M_\odot) < 11.5$. Note that when using the central velocity dispersion as a tracer of the bulge potential we recover the same overall trends as for pure photometric tracers like S\'ersic index and concentration, thus confirming that our results are not dominated by the uncertainty in the photometry. 

\section{Kinematic misalignment}
\label{sec5}

Kinematic decoupling between the gaseous and stellar components has been observed to be common in local ETGs \citep{Bertola1999, Sarzi2006, Davis2011}, with hints of this behaviour being already present at high redshift \citep{Wisnioski2015}. The distribution of misalignments between the gas and stars is a direct probe of the origin of the gas in ETG. 
Theoretically, a newly formed gas disc is predicted to precess due to the effect of the
gravitational potential of the stars, with gas precessing faster at smaller radii, where the
torque is larger. Cloud-cloud collisions will also work to realign the differently precessing
gas rings, and the system will eventually settle into a configuration where the gas and stellar
components are either aligned or misaligned by 180$^{\circ}$. The relaxation time for this
process was initially predicted to be of the order of the dynamical time \citep{Tohline1982,
Lake1983}, however recent work based on hydrodynamics in the cosmological framework has
highlighted that misalignments might persist on much longer (several Gyr) time-scales
\citep{VandeVoort2015}, mostly because of the effect of continuous cosmological accretion.
This longer relaxation time-scale also provides a better fit to the misalignment distributions observed in the literature \citep{Davis2015}. For weakly triaxial systems stable gaseous orbits are allowed both in the plane containing the long and intermediate axes and the plane of the short and intermediate axis \citep{Franx1991}. If the origin of the gas is external, therefore, the distribution of the gas-stars kinematic misalignment should display three peaks, respectively at $0, 90, 180^\circ$, with the peak around $90^\circ$ being weaker and scaling with the degree of triaxiality.

In this section we derive the kinematic position angle (PA) of both gas and stars following the procedure in \cite{Krajnovic2006}. In short, having fixed the kinematical centre to the photometric one, the algorithm constructs a bi-antisymmetric version of the velocity field and compares it with the observed velocity field. The resulting PA is thus only representative of the axis of symmetry of the velocity field and the measurement does not imply that the galaxy is best fitted with a thin disc model. In our sample four galaxies present particularly complex gaseous kinematics with no obvious symmetry axis and have been excluded from the following analysis. Two of these galaxies show approaching/trailing features while other two show evidence for extreme changes in the direction of rotation of the gas (MaNGA-ID {\tt 1-38167}, {\tt 1-274663}). 

The remaining eLIER galaxies in our sample display a wide range of gas velocity field patterns. While some galaxies are consistent with a gaseous disc rotation, in other systems the velocity field appears disturbed. Overall, the large majority of galaxies in this class show coherent velocity shear patterns, showing both red and blue-shifted emission. However, these gas flows are not always associated with large-scale rotation. For example, in \cite{Cheung2016} we propose a biconical outflow model for an eLIER galaxy (MaNGA-ID {\tt 1-217022}) that has recently been subject to a minor merger event (as demonstrated by the presence of a nearby companion and the detection of neutral gas via the NaD absorption feature).

\begin{figure} 
\includegraphics[width=0.49\textwidth, trim=80 80 80 120, clip]{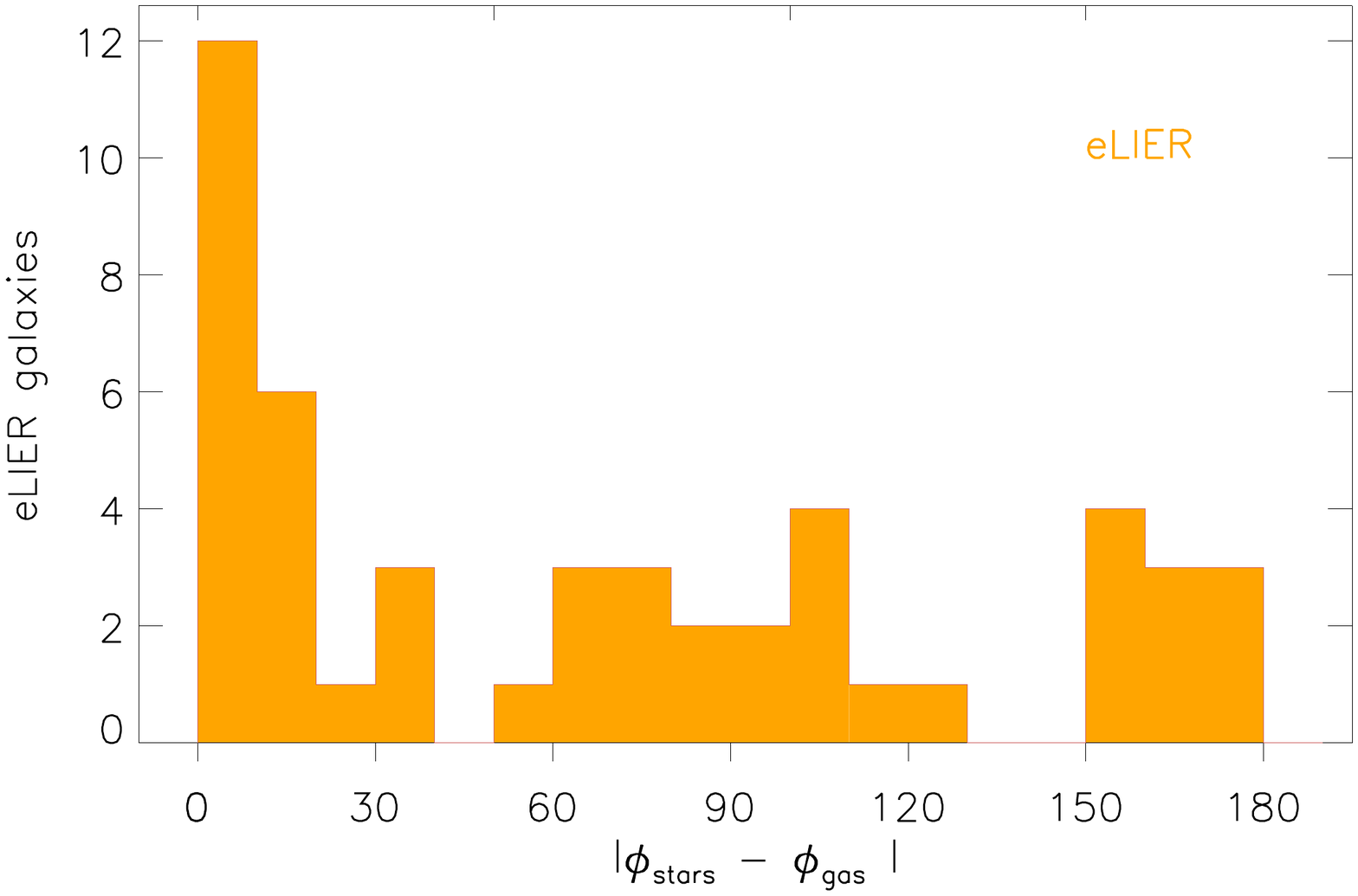}
\includegraphics[width=0.49\textwidth, trim=80 80 80 150, clip]{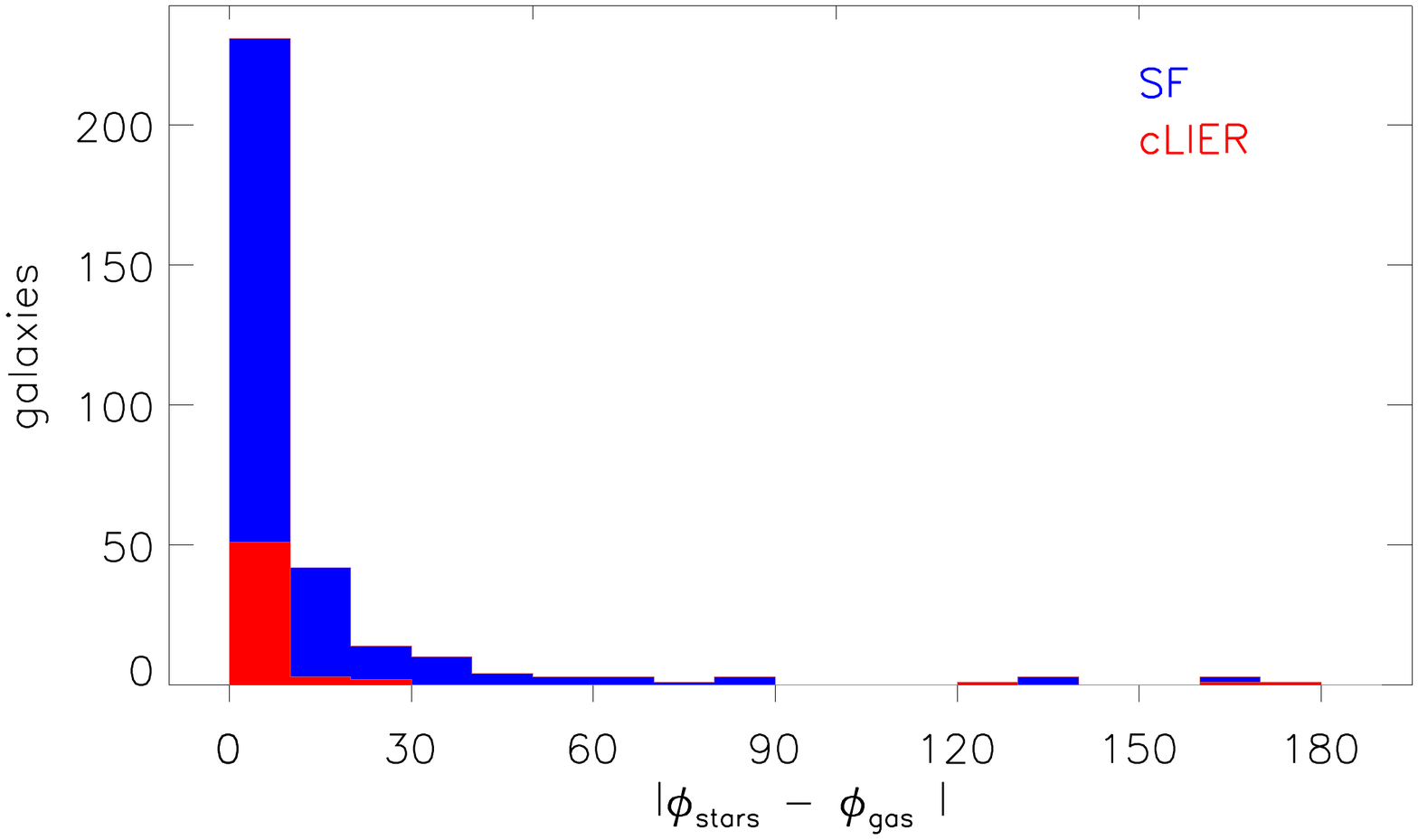}
\caption{Histograms of the distribution of kinematic misalignments between the position angle of the major axis of the stellar component and of the ionised gas for the eLIER galaxies (top) and SF and cLIER galaxies (bottom) in the current MaNGA sample.}
\label{fig5.1}
\end{figure}

The observed distribution of stars-gas misalignment for the eLIER galaxies in our sample is shown in Fig. \ref{fig5.1}, with 30 out of 49 galaxies having misalignments larger than 30$^{\circ}$. Making the appropriate volume corrections, necessary in order to take into account the MaNGA selection function, we infer that $65 \pm 7 \%$ of eLIERs are misaligned with $\rm |\phi_{stars}-\phi_{gas}| > 30 ^\circ$. This number is significantly larger from that inferred by the SAURON and ATLAS$\rm^{3D}$ surveys \citep{Sarzi2006, Davis2011}. However, both surveys adopt a pure morphological selection, thus including galaxies with residual star formation into their sample, which would not be classified as eLIERs. Galaxies with detectable star formation (SF and cLIERs) have a misalignment distribution strongly peaked at zero (Fig. \ref{fig5.1}, bottom panel, only $11 \pm 2 \%$ with  $\rm |\phi_{stars}-\phi_{gas}| > 30 ^\circ$), thus a small contamination from SF or cLIER galaxies can bring our result in line with previous work. Moreover our results confirm the theoretical prediction for the case of external accretion, as the observed distribution of misalignments for eLIERs is peaked at $0, 90$ and $180^\circ$. 

In absence of internal processes (i.e. stellar mass-loss), the misalignment histogram is
predicted to be {\it symmetric} around 90$^\circ$ \citep{Davis2015}. Making the assumption of
isotropic accretion and long relaxation times, one would naively expect 150/180$\sim$83\% of
galaxies to have misalignments larger than  $30^\circ$. The presence of a stronger peak at zero
and the fact that the number of misaligned galaxies is lower than a naive estimate for
isotropic accretion implies that internal processes are also likely to play a role, although
secondary, in shaping the observed misalignment distribution. In particular, stellar mass-loss
will not only inject into the ISM some amount of gas sharing the same kinematic properties
as the parent stars, but would also create additional torque on the accreted misaligned gas and contribute to realign it with the stellar kinematic field. Both effects
would contribute to increase the fraction of aligned galaxies.

\section{The effect of environment}
\label{sec6}

\begin{figure*} 
\includegraphics[width=0.49\textwidth, trim=50 40 50 50, clip]{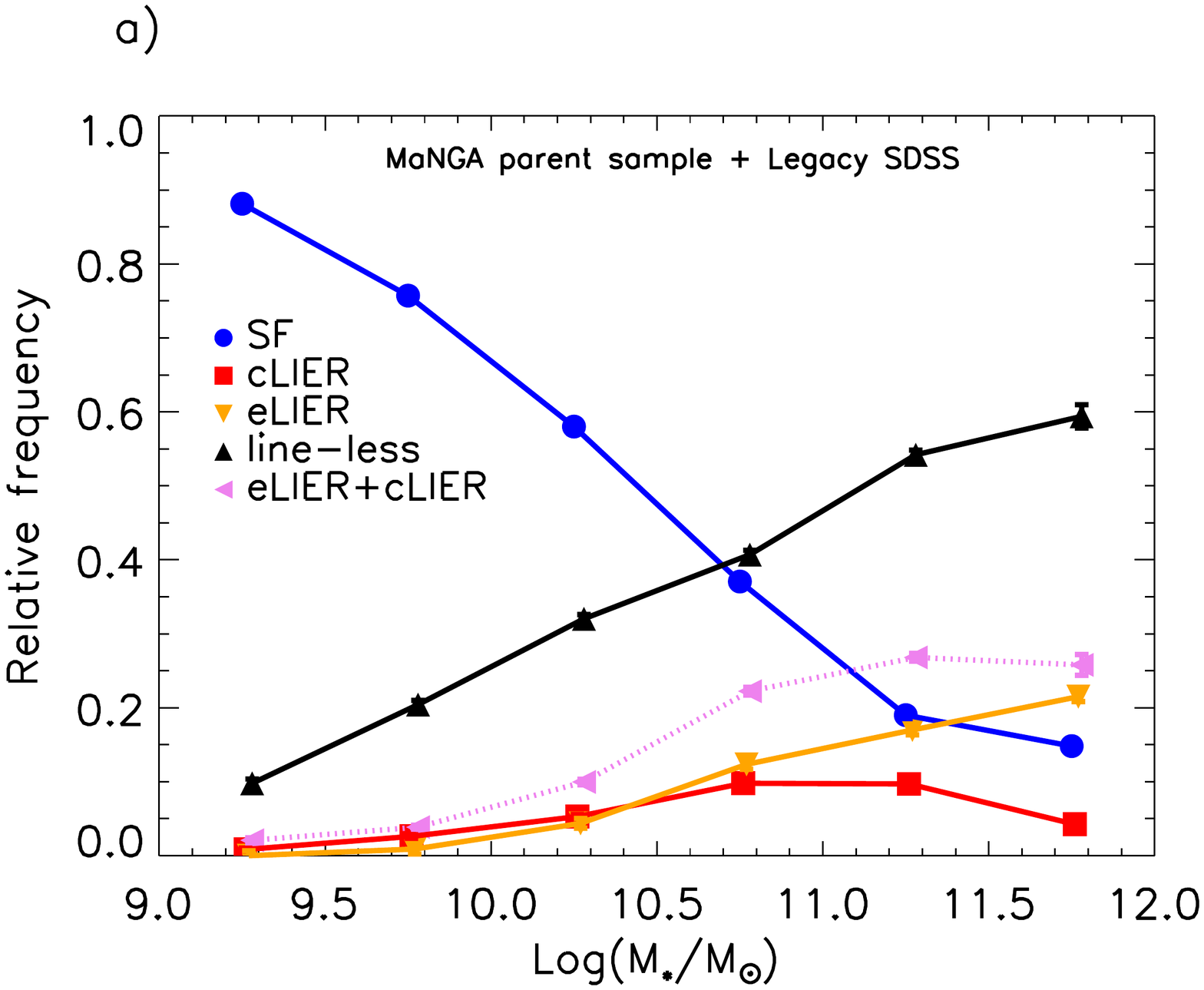}
\includegraphics[width=0.49\textwidth, trim=50 40 50 50, clip]{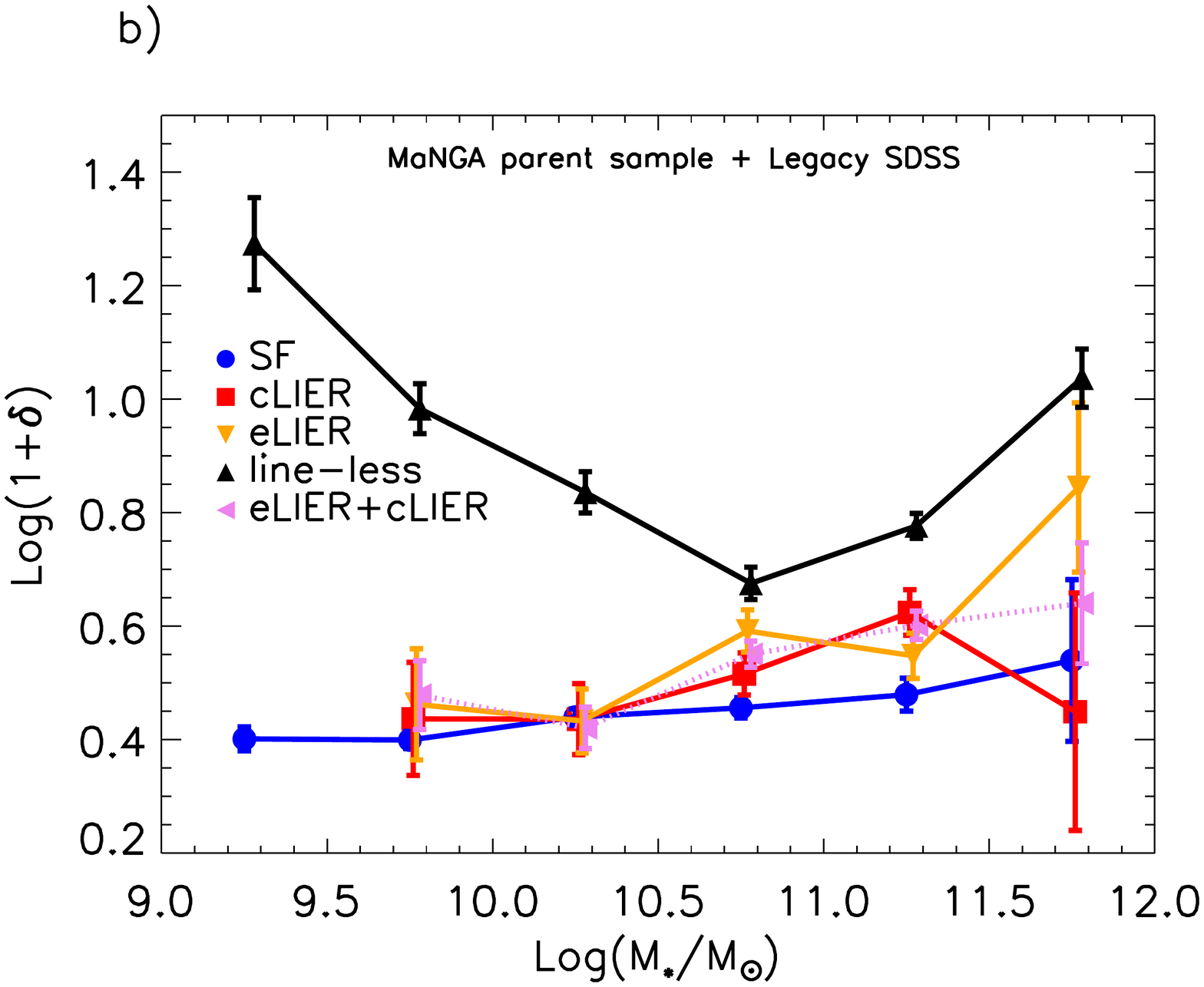}
\includegraphics[width=0.49\textwidth, trim=50 40 50 50, clip]{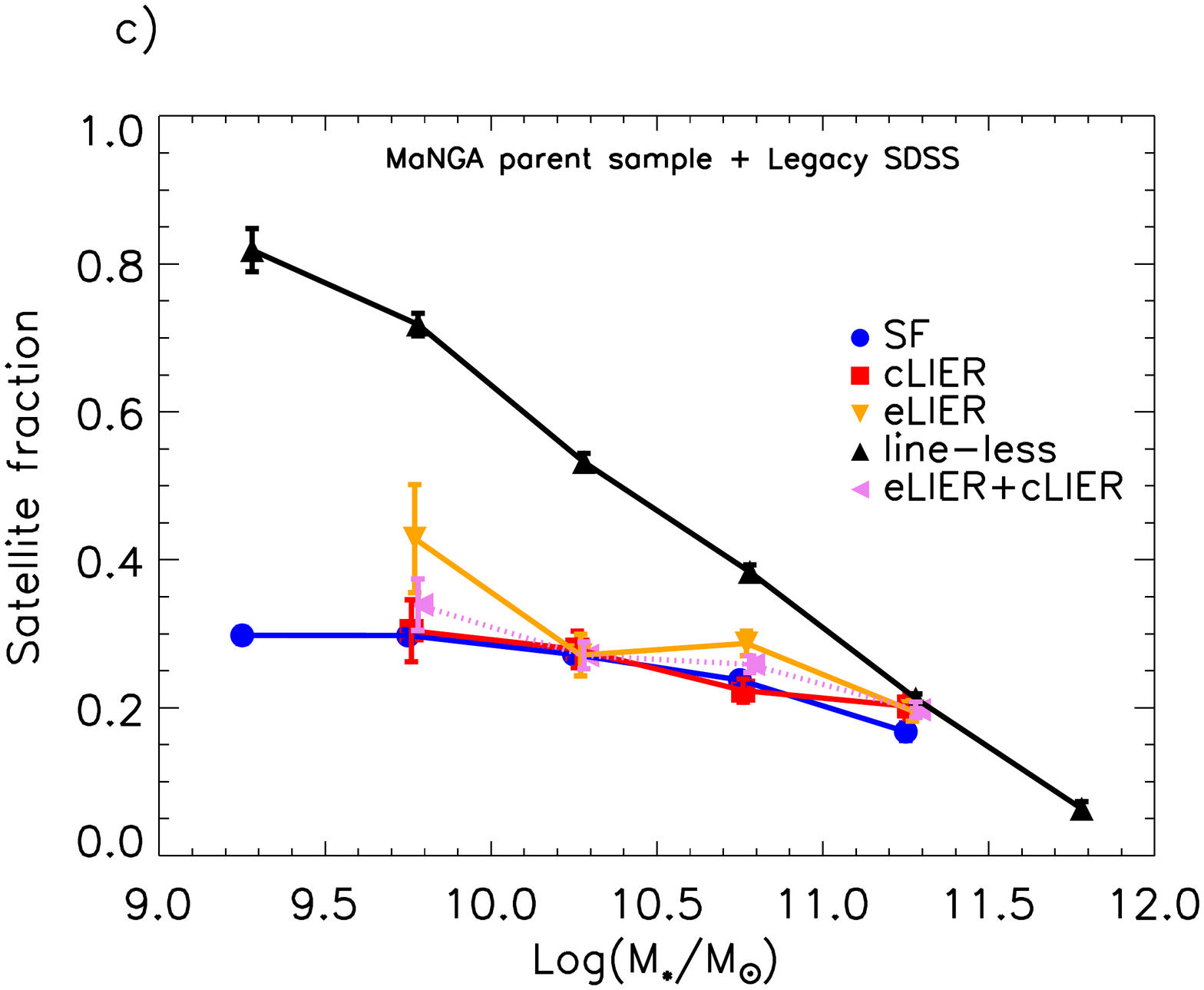}
\includegraphics[width=0.49\textwidth, trim=40 40 50 50, clip]{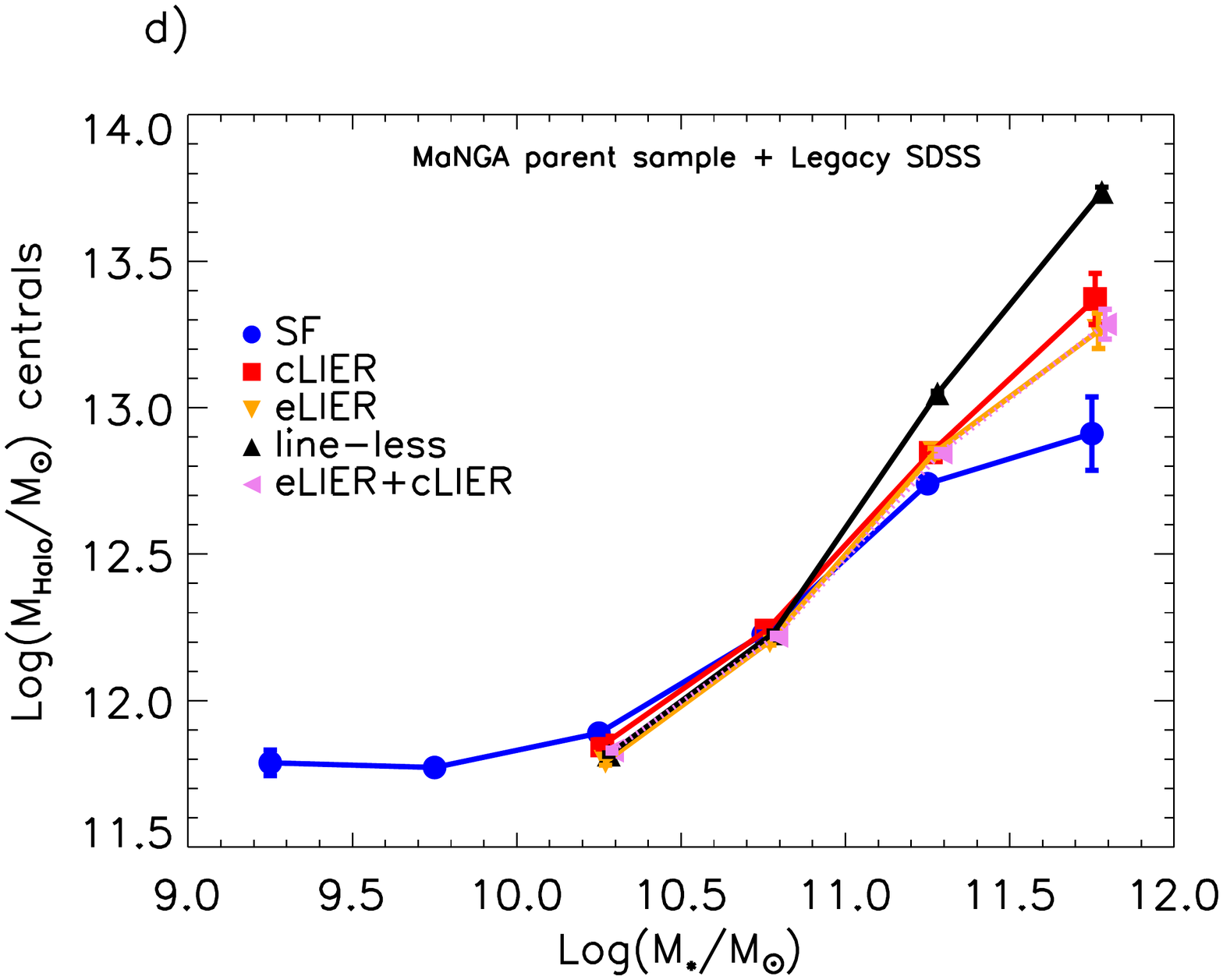}

\caption{\textit{(a)} The relative frequency of galaxies in different emission-line-based classes derived from the MaNGA parent catalogue and legacy SDSS spectroscopy in bins of stellar mass. \textit{(b)} Average overdensity as a function of stellar mass for galaxies of different emission-line classes, based on the MaNGA parent catalogue and SDSS legacy spectroscopy. \textit{(c)} Average fraction of galaxies classified as satellites in the \protect\cite{Yang2007} group catalogue for each emission-line based galaxy class as a function of stellar mass. The galaxy sample is taken from the MaNGA parent sample, complemented with SDSS legacy spectroscopy. \textit{(d)} Average halo mass for galaxies classified as central in the \protect\cite{Yang2007} group catalogue for each emission-line bases galaxy class as a function of stellar mass. The galaxy sample is taken from the MaNGA parent sample, complemented with SDSS legacy spectroscopy. In each plot, the error bar represents the error on the mean and only bins containing more than four galaxies are shown. The dotted line (eLIER+cLIER) refers to the union of both cLIER and eLIER classes.}
\label{fig6}
\end{figure*}

Fundamental properties of the bimodal galaxy population, including colour, SFR and passive fraction are found to depend both on stellar mass and environment \citep{Blanton2005, Baldry2006, Peng2010}. Using data from legacy SDSS, it has been demonstrated that the change in EW(H$\alpha$) distribution as a function of environment is predominantly due to a change in the relative number of SF and passive galaxies \citep{Balogh2004}. If one restricts the study to SF galaxies, the EW(H$\alpha$) (or equivalently the specific SFR) is observed \textit{not} to depend on environment, i.e. if a galaxy forms stars on the main sequence, environments does not effect its SFR \citep{Peng2010}, which is consistent with the finding that stellar population ages do not depend on environment \citep{Thomas2010}. However, passive galaxies with measurable EW of line emission (eLIERs in our classification) have long been associated with inefficiently accreting AGN \citep[e.g.][]{Kauffmann2003a, Kewley2008} and their role as red sequence galaxies with residual gas has not yet seen suitably discussed in the context of environmental studies. 

In this work we have highlighted the similarities of eLIER to line-less galaxies in terms of the stellar populations and morphological properties. Given these observed similarities, it is logical to ask whether environment plays a role in the presence or absence of line emission on the red sequence. Some evidence in this direction was reported by \cite{Sarzi2006}, who found a small difference in the fraction of detected line emission between ETGs in the Virgo cluster compared to the field.

In order to address this question with a large sample of galaxies, in this section we make use of the MaNGA parent galaxy sample (Wake et al., \textit{in prep.}), and study the emerging trends as a function of different environmental measures. We do not use the MaNGA data directly because the sample of LIER galaxies currently available is too small to reliably study environmental trends in several stellar mass bins. The MaNGA parent sample, on the other hand, includes all galaxies from the NSA catalogue that match the MaNGA target selection criteria ($\sim$ 30000 galaxies). We use legacy SDSS spectroscopy \citep{Strauss2002}, and make use of the line flux measurements from the MPA-JHU catalogue \citep{Kauffmann2003a, Brinchmann2004, Tremonti2004} to compute the BPT classifications of each galaxy. In order to study trends as a function of stellar mass we correct for the effect of the MaNGA selection function.

In order to adapt the cLIER/eLIER distinction to single-fibre legacy SDSS, which does not cover the outer regions of galaxies, we make use of the integrated UV-optical colours, based on the findings in section \ref{sec4.1} above. We assume that both cLIERs and eLIERs would appear as LIER galaxies in single fibre spectroscopy. A galaxy appearing as LIER in single-fibre spectroscopy is classified as cLIER if it lies on the blue cloud or the green valley ($NUV - r \rm< 5$) and eLIER if it lies on the red sequence ($NUV - r \rm> 5$) using UV-optical colours. We make use of the MaNGA parent sample instead of the full legacy SDSS in an attempt to minimise the role of aperture effects in this classification, an effect that is otherwise difficult to quantify. According to section \ref{sec4.1}, our mixed spectroscopic + photometric eLIER/cLIER distinction introduces a 19\% contamination for eLIERs and 13\% contamination for cLIERs. Similarly, in the attempt to replicate in the best possible way the MaNGA-based classification, galaxies with EW(H$\alpha$) $<$ 1.0 \AA\ within the 3$''$ SDSS fibre are considered line-less. Seyfert AGN, as identified using the [SII] BPT diagram, are excluded from this analysis.

In Fig. \ref{fig6}a) we show the fraction of galaxies in each class as a function of mass in the MaNGA parent sample. This figure is to be compared with the equivalent figure in \textit{Paper I} (Fig. 6) for the MaNGA sample which is also used in this work. Qualitatively the mass trends for each class are similar, however some differences are also present. In particular, the classification based on legacy SDSS finds less LIER galaxies than in the MaNGA data, which might be due to residual aperture effects.

In order to study environmental effect, we make use of adaptive aperture environmental measure presented in \cite{Etherington2015}. The cosmic density $\rho$ field is calculated using a cylinder centred on each galaxy and with length of 1000 $\rm ~km~s^{-1}$ and radius given by the distance to the fifth nearest neighbour more luminous than $\rm M_{r} < -20.26$ (where $\rm M_{r}$ is the k-corrected r-band magnitude). The environment estimator is cast in terms of an overdensity ($\delta = (\rho - \rho_m)/\rho_m$) relative to the average density of tracer galaxies $\rho_m$ and is most sensitive to scales around 2 Mpc \citep{Etherington2015}. 

We further make use of the environmental information provided by the \cite{Yang2005, Yang2007} group catalogue. The catalogue is based on a halo-finding algorithm applied to the NYU-Value Added Catalogue \citep{Blanton2005}, based on legacy SDSS DR7 data. Galaxies belonging to the same dark matter halo are defined to constitute a `group' (where the term group is used here irrespective of the number of galaxies in the association). The most luminous galaxy in each group is defined to be the `central', while the other galaxies in each group are defined to be `satellites'. Note that, following this definition, centrals do not need to necessarily lie at the geometric centre of the dark matter halo. Several galaxies are assigned to haloes that contain no other galaxy within them; these galaxies are called centrals even though no satellites are detected in SDSS (although deeper observations might reveal their presence). The halo mass ($\rm M_{halo}$) is calculated for each group by \cite{Yang2007} considering the characteristic integrated luminosity (above some flux limit, and corrected for incompleteness) of the group members calibrated against the dark matter masses derived from models.

In Fig. \ref{fig6}b) we show the mean environmental density as a function of stellar mass for each galaxy class (SF, cLIER, eLIER and line-less). The plot demonstrates that (1) At fixed stellar mass, line-less galaxies always live in denser environments than eLIER galaxies. The relative difference in environmental density between the two galaxy classes is large at both the lowest and highest stellar masses, with a minimum difference at $\rm  M_\star \sim 10^{10.5}~M_\odot$. (2) There is only a marginal difference in environmental density between SF galaxies and cLIER or eLIERs, with eLIERs living in marginally denser environments. 

For the purpose of the study of line emission in the overall galaxy population, the most
striking feature of of Fig. \ref{fig6}b) is the large difference in environmental density
between eLIERs (i.e. red sequence galaxies with line emission) and line-less galaxies (i.e. red
sequence galaxies with EW(H$\alpha) <$ 1.0 \AA). \textit{The legacy SDSS data demonstrate
clearly that the lack of line emission on the red sequence is associated with higher
environmental densities.} We discuss the possible physical processes that could be responsible
for this phenomenon in section \ref{sec7.2}.

Fig. \ref{fig6}c) demonstrates that line-less galaxies have a substantially higher satellite fraction than any other galaxies class at all stellar masses lower than $\rm M_\star \sim 10^{11}~M_\odot$. Hence, the increase in the environmental density of line-less galaxies at low stellar masses is mostly driven by the fact that these galaxies are preferentially satellites. 

At high stellar masses, on the other hand, line-less central galaxies have a higher mean halo mass than other galaxy classes (Fig. \ref{fig6}d). Higher mass haloes tend to have a higher number of satellites and also to be more strongly clustered, thus explaining the increase in environmental density at the high-mass end. The figure also shows that in the highest mass bin, SF galaxies reside in lower-mass haloes than LIER galaxies (both eLIERs and cLIERs).


\section{Discussion: an evolutionary scenario for LIERs}
\label{sec7}

\subsection{Inside-out quenching and the nature of cLIER galaxies}
\label{sec7.1}

\subsubsection{A slow quenching pathway for late-type galaxies}

The properties of cLIERs are such that they can be considered objects in transition (in the process of being `quenched') from the SF blue cloud towards the quiescent red sequence. The MaNGA spatially resolved data further demonstrate that star formation in these galaxies terminates from the inside out, with LIER emission appearing in the central regions, while star formation continues in the disc. The quenching process is not only evident via the lack of detected star formation in diagnostic diagrams, but also in the stellar population properties, as LIER emission is strongly correlated with old stellar populations (section \ref{sec3.3}, also \textit{Paper I}, section 7).

The process leading to the quenching of the central regions is likely to be gentle, since it does not involve a change in the overall morphology and kinematics of the galaxy, although it is associated with the presence of a large central spheroidal component. cLIER galaxies are thus consistent with the idea of a `slow quenching mode' for LTGs \citep[e.g.][]{Schawinski2014, Smethurst2015}. 
Different lines of evidence point towards the need of a such a `slow quenching' pathway for LTGs associated with the increasing importance of the bulge component. \cite{Cappellari2013}, for example, argue that the quenching process involved in the production of fast rotator ETGs needs to increase the bulge size, while at the same time removing the gas reservoir and/or shutting down star formation and maintaining the flatness of the disc at large radii. At higher redshift \cite{Bell2012} and \cite{Bundy2010} argue for a similar slow quenching pathway initiated by gradual changes in internal structure.

A possible quenching mechanism, much discussed in the literature \citep[e.g.][]{Wake2012, Bluck2014}, assumes the higher central spheroidal masses to be directly connected to the cause of galaxy starvation via AGN radio-mode feedback. Indeed, larger bulge/spheroid masses imply larger black hole masses, hence higher capability of halo heating through radio-mode episodes \citep{Heckman2014}, which in turn may result into lack of gas cooling on to the host galaxy. While the inclusion of AGN radio-mode feedback in galaxy formation models has proven to be able to successfully solve a number of longstanding problems, direct evidence for this feedback mode in galaxies with $\rm M_\star \sim 10^{10} ~ M_\odot$ remains lacking. It may be natural to assume that this feedback mode, generally predicted to be important in much more massive haloes (clusters, e.g. \citealt{Fabian2012}) would equally affect lower mass haloes. However, other alternatives sharing the same scaling relation cannot be ruled out.

The interruption of the inflow of cold gas on to the galaxy (`strangulation'), due to either heating of the gas in the halo or via halo mass shock heating, can lead to an inside-out quenching signature, assuming that the remaining cold gas is consumed more quickly in the central regions of galaxies. As illustrated by \cite{Leroy2008} the efficiency of star formation (i.e. the number of stars formed per unit gas mass) is not constant across galaxy discs, but is observed to steadily decrease with galactocentric distance, thus making this scenario a viable option. 

\subsubsection{Secular processes and bulge growth}

While a gas exhaustion scenario naturally produces slow, inside-out quenching, the simple strangulation scenario does not directly address the larger bulge components observed in cLIER galaxies compared to SF galaxies of the same stellar mass (unless one makes the additional assumption that bulge mass is driving the heating of the halo via radio-mode feedback).


Alternatively, the larger bulge masses, together with the implied slow time-scale for the quenching process, may argue in favour of secular processes (i.e. mechanisms internal to the galaxies) playing a role in the quenching process in cLIERs. Bars, for example, tend to funnel gas towards the galactic central regions, thus contributing to the formation of (pseudo)bulges \citep{Kormendy2010, Simmons2013, Cheung2013}. Moreover, simulations show that the torques and stresses created by both bars and bulges on the cold gas component can lead to lower star formation efficiency, especially in the regions swept up by the bar \citep{Emsellem2014, Fanali2015}. Indeed, bar fraction is known to correlate with the galaxy's stellar mass, colour and specific SFR \citep{Nair2010, Masters2011}, in the sense of redder and more massive, lower specific SFR discs having higher bar fraction. Moreover, bar fraction is observed to be higher in H\textsc{i} poor galaxies \citep{Masters2012}, although no similar investigation of the molecular gas content of a large sample of barred and unbarred galaxies has been possible to date (see \citealt{Saintonge2012} for the largest sample to date). 

Using the current MaNGA sample, for galaxies with $\rm log(M_\star/M_\odot) > 10$ the bar fraction (defined as disc galaxies with $\rm p_{bar}>0.3$ from Galaxy Zoo 2) is $32\pm5$\% for the SF discs, and $41\pm11$\% for the cLIERs. We thus conclude that the difference in bar fraction between the two classes is not significant with the current sample size, and a larger sample is needed to observationally test the relative occurrence of bars in SF galaxies are cLIERs.

Models predict that disc instabilities represent the dominant contribution to the formation of bulges in massive galaxies ($\rm M_\star > 10^{10}~M_\odot$, \citealt{DeLucia2011}). Disc stability can be parametrised in terms of the stability parameter (Q, \citealt{Toomre1964}), which in the case of gas+stars systems can be extended to include the two components \citep{Jog1984, Elmegreen1995, Martig2009, Romeo2011}. The presence of a significant bulge goes towards increasing the stability of a cold gas disc to gravitational collapse. Indeed, results from the COLD-GASS survey \citep{Saintonge2011} demonstrate that galaxies with significant bulges have lower star formation efficiencies \citep{Saintonge2012}. Spatially resolved observations of the gas component for the MaNGA sample would be crucial in testing the role of the bulge in driving the star formation efficiency in different galaxy classes.

Star formation efficiency for cold gas is found to be systematically lower in ellipticals \citep{Martig2013, Davis2014} and in the bulge of our own Galaxy \citep{Longmore2013}. Therefore, the absence of the star formation in the central regions of cLIERs could be due to a significantly lower star formation efficiency, even in presence of a large molecular gas reservoir, in a similar fashion as the `central molecular zone' in the Milky Way, which is found to have a star formation efficiency approximately 10 times lower than the Milky Way disc. In fact, the Milky Way itself is a likely cLIER candidate, given its inferred green valley optical colours and SFR lower than the star formation main sequence \citep{Mutch2011}.

In this context, secular evolution processes (interaction with a bar or spiral density waves) and radial flows represent a possible explanation for the presence of cold gas in the central regions of cLIERs, which is required in order to explain the observed LIER emission. The kinematic correspondence between the stars and gas, which is preserved in the central regions of cLIER galaxies, further supports a scenario where the gas in the central region of cLIER galaxies is either integral part of the disc or has been acquired from it. Indeed, the absence of line-less bulges in the current sample is a strong indication that central regions of bulge + disc systems are always more gas rich than galaxies on the red sequence and points towards the importance of the SF disc as a possible reservoir of gas in these systems.

In summary, cLIER galaxies are consistent with both a scenario of slow gas exhaustion and a scenario of reduced star formation efficiency in the bulge, driven by dynamical processes. The cold gas content of the central regions of these galaxies is not known, however some amount of gas must be present in order to allow for the observed LIER emission. This gas shares the same kinematics of the SF disc and is thus possibly driven into the central regions by bars or other disc instabilities. Followup resolved studies of the cold gas content of bulge + disc systems are needed in order to shed further light into the reasons for the absence of star formation in the central regions of cLIERs.

\subsection{Ionised gas on the red sequence and the nature of eLIER galaxies}
\label{sec7.2}

\subsubsection{Origin of the gas in red galaxies}
It has long been argued that red sequence galaxies have a large and plentiful source of ionised gas in their old stellar populations. Stellar evolution predicts that an old stellar population returns of the order of half its stellar mass to the ISM over a Hubble time \citep{Ciotti1991, Padovani1993, Jungwiert2001}. Mass return is dominated by stellar mass-loss during the evolution of intermediate and low mass stars along the red giant branch and the post AGB phases. \cite{Padovani1993} predict that, for a Salpeter initial mass function, a 15 Gyr old population will lead to a mass-loss rate per unit stellar mass of 

\begin{equation}
\rm \dot{M}_\star = 2.3 \cdot 10^{-12} ~M_\star~ yr^{-1} .
\end{equation}

The mass of ionised gas can be estimated from the H$\alpha$ luminosity as 

\begin{equation}
\rm M_{HII} = 2.3 \cdot 10^3 ~ \left( \frac{L_{H\alpha}}{10^{38}~erg~s^{-1} } \right) \left( \frac{10^2~cm^{-3}}{n_e} \right) ~M_{\odot}.
\end{equation}

For typical integrated H$\alpha$ luminosities observed in eLIERs, this implies a mass of about $\rm \sim 10^4-10^5~M_{\odot}$. Therefore, for eLIER galaxies of typical mass of about $\rm 10^{10.5}~M_{\odot}$, the typical nebular line luminosities can easily be accounted for by gas injected into the ISM by stellar mass-losses over a time-scale of only a few million years.

However, the kinematic signatures of the ionised gas are expected to be similar to those
of the parent stars. This predictions is at odds with the observational evidence that a large
fraction ($\sim$ 65 \%) of eLIER galaxies are kinematically misaligned. Thus, stellar mass
loss can only be responsible for a minor fraction of the gas associated with
LIER emission. Instead it is likely that the gas supplied by stellar mass-loss is quickly transferred into the hot X-ray emitting phase, with only a small fraction succeeding to cool. Indeed, the standard paradigm of the ISM in red sequence galaxies assumes that supersonic motion of the ejecta through the ambient hot medium generates a strong shock that thermalises the gas to the kinetic temperature of the stars, $\sim 10^{6} - 10^{7}$ K, on a time-scale of $\sim 10^5 - 10^6$ yr \citep{Sanders1981, Mathews1990}. Episodic AGN activity could also be responsible for balancing cooling \citep{Ciotti2007, Schawinski2007, Cheung2016} and maintaining the gas in a hot state.

In summary, the existence of a large population of red sequence galaxies with $\rm EW(H\alpha) < 1 \AA$ and the same stellar population properties as eLIERs suggests that stellar mass-loss cannot be the main source of cool gas in these galaxies. This inference is confirmed by the fact that in our sample of eLIERs external accretion is required in at least 65 \% of the population to explain the observed kinematic misalignment between gas and stars.

\subsubsection{The role of environment on the red sequence}

As discussed in section \ref{sec6}, environmental effects are known to play a key role in determining the passive fraction of galaxies \citep{Dressler1980, Peng2010}. In this work, we have shown that, on top of determining the passive faction, environmental effects also affect the presence of line emission and thus the cold gas content in red sequence galaxies. In particular, we have demonstrated that red sequence galaxies with small or undetected line emission (EW(H$\alpha$) $<$ 1 \AA) live in denser environments, both as centrals and satellites, than eLIERs of the same mass. 

At the high end of the stellar mass function line-less galaxies are mostly the central galaxies in high-mass haloes. Their position within a large halo would both impede accretion of external cold gas and heat up the gas recycled via stellar evolution, thus leaving the galaxy line-less. Analysis of the stellar kinematics of the MaNGA data further demonstrates that at high masses most galaxies classified as slow rotators are line-less, as expected from the kinematics-density relation \citep{Cappellari2011b}. These central, massive, slowly-rotating line-less galaxies represent the striking endpoint of galaxy evolution.

At the low mass end, environment is known to play a crucial role in making galaxies line-less,
albeit via a different route \citep[e.g.][]{Geha2012}. We demonstrate in Fig. \ref{fig6} that
for $\rm M_\star < 10^{10}~M_\odot$ more than 70 \% of line-less galaxies are satellites, in
contrast with the much lower satellite fraction for eLIERs ($\sim$ 40 \%). Satellites are
generally not thought to accrete gas after falling into the hot environment of
galaxy overdensities (or anyhow to accrete at a substantially lower rate than centrals, \citealt{Keres2009}) and are likely to undergo some amount of ram-pressure stripping as they enter a massive halo. Moreover, while at first the merger rate is enhanced as galaxies fall into rich haloes, once the galaxies virialise the large velocity dispersion within cluster considerably suppresses the subsequent merger rate \citep{vanDokkum1999}, thus making it harder to obtain gas from gas rich mergers. While it is possible that some amount of dense gas survives the infall event and allows the satellite galaxy to continue forming stars for a time-scale comparable to the depletion time \citep{Wetzel2013}, once star formation ceases the strong environmental effects are likely to seriously disrupt the diffuse gas component and finally leave the galaxy line-less.


\section{Conclusions}
\label{concl}

In this paper we study the properties of galaxies based on their spatially resolved ionised gas emission. We make use of spatially resolved spectroscopy for a sample of 586 galaxies from SDSS-IV MaNGA, and classify galaxies according to the classification scheme introduced in \cite{Belfiore2016a}, which properly accounts for the ubiquitous presence of low ionisation emission line regions (LIERs). Within this framework non-active galaxies are classified in four classes: (1) star forming galaxies (SF), (2) galaxies characterised by LIER emission at small galactocentric radii and star formation at larger radii (cLIERs), (3) galaxies characterised by LIER emission at all radii where line emission is detected (eLIERs), (4) Line-less galaxies, operationally defined to have average EW(H$\alpha$) $<$ 1 \AA\ within 1.0 $\rm R_e$. 

In this work we place these galaxy classes within the context of the evolving galaxy population by studying their integrated colours and SFR, morphology and kinematics. We complement the MaNGA data with environmental information for the much larger MaNGA parent sample ($\sim$ 30~000 galaxies with legacy SDSS spectroscopy) to be able to disentangle the effect of mass and environment on line emission on the red sequence. In the following we summarise the fundamental properties of the two new galaxy classes we have focused on.

\paragraph*{Central LIER galaxies (cLIERs)}
\begin{enumerate}
\item{Stellar populations: cLIERs span a wide range of stellar ages, with old central LIER regions and outer regions characterised by young stellar populations.}
\item{Colours and SFR: cLIERs lie preferentially in and around the green valley in UV-optical colours, although they appear increasingly consistent with the red sequence using optical colours alone ($u-r$ or $g-r$). In the SFR-$\rm M_\star$ diagram cLIERs lie 0.8 dex below the main sequence of star forming galaxies, albeit with a large scatter.}
\item{Morphology: cLIER galaxies are visually classified as disc galaxies, but have larger bulge fraction than star forming galaxies of the same mass, as traced via photometric indices and spectroscopic stellar velocity dispersion measurements.}
\item{Kinematics: cLIERs display regular disc-like kinematics in both gas and stars, which are mutually aligned.}
\item{Environment: cLIERs live in slightly denser environments than SF galaxies for $\rm log(M_\star/M_\odot) = 10.5- 11.5$, but have the same satellite fraction.}

\end{enumerate}
cLIERs are consistent with originating from SF galaxies that are slowly quenching. The inside-out quenching pattern is consistent with both gas exhaustion and/or a decrease in the star formation efficiency. In either case, the presence of a bulge is closely connected to the lack of central star formation in these galaxies. The star forming disc is the likely source of the LIER emitting gas present in the central regions.

\paragraph*{Extended LIER galaxies (eLIERs)}
\begin{enumerate}
\item{Stellar populations: eLIERs have old stellar populations, indistinguishable from those of line-less galaxies.}
\item{Colours and SFR: eLIERs lie on the red sequence in colour-magnitude diagrams, with only a small contamination to the green valley.}
\item{Morphology: eLIER galaxies are visually classified as early type galaxies and have similar bulges to those of line-less galaxies of the same stellar mass.}
\item{Kinematics: eLIERs display a variety of gas kinematics, but generally present well-defined rotation/flows on kpc scales. The velocity field of gas and stars are often observed to be misaligned ($65 \pm 7  \%$  of eLIERs are misaligned by more than 30$^\circ$).}
\item{Environment: eLIERs live in denser environments than SF and cLIERs at high masses ($\rm M_\star > 10^{10.5} M_\odot$), as expected from the well-known correlation between colour and passive fraction. More importantly eLIERs tend to live in less dense environments than line-less galaxies of the same mass. At the high mass end, line-less galaxies are mostly centrals in high mass haloes, while at the low mass end line-less galaxies are mostly satellites.}

\end{enumerate}
eLIERs are red sequence galaxies with residual cold gas, acquired mostly via external sources. While this gas does not form stars, it is illuminated by the diffuse ionisation field from hot old stars and shines with LIER line ratios. LIER emission is suppressed in high density environments, likely because of lack of gas, which has either been consumed due to the shut down of inflows or otherwise heated and/or tidally stripped by the hot halo gas.

%

\section*{Acknowledgements}
\begin{small}
We are grateful to the referee, A. Renzini, for his comments and insight that have greatly improved the quality of the paper.
F.B., R.M and K.M acknowledge funding from the United Kingdom Science and Technology Facilities Council (STFC). R.M. acknowledges support from the European Research Council (ERC) Advanced Grant 695671 `QUENCH'. A.R-L acknowledges partial support from the DIULS regular project PR15143. M.B was supported by NSF/AST-1517006. K.B. was supported by World Premier International Research Centre Initiative (WPI Initiative), MEXT, Japan and by JSPS KAKENHI Grant Number 15K17603. A.W. acknowledges support from a Leverhulme Early Career Fellowship. A.D. acknowledges support from The Grainger Foundation. The authors are thankful to Y. Peng and D. Goddard for their help with the environmental measures for the SDSS galaxy sample and to M. Blanton for developing and maintaining the NASA-Sloan Atlas; to the members of the SDSS-IV MaNGA collaboration, in particular the dedicated team of observers at APO. The visual classification of the Galaxy Zoo galaxies was made by more than 100~000 volunteers. Their contributions are acknowledged at {\tt http://www.galaxyzoo.org/Volunteers.aspx}. This work makes use of data from SDSS-I-II and IV. This research made use of {\tt Marvin} (Cherinka et al., in prep), a core Python package and web framework for MaNGA data, developed by Brian Cherinka, Jos\'e S\'anchez-Gallego and Brett Andrews.

Funding for SDSS-I-II and SDSS-IV has been provided by the Alfred P.~Sloan Foundation and Participating Institutions. Additional funding for SDSS-II comes the National Science Foundation, the U.S. Department of Energy, the National Aeronautics and Space Administration, the Japanese Monbukagakusho, the Max Planck Society, and the Higher Education Funding Council for England.  Additional funding towards SDSS-IV has been provided by the U.S. Department of Energy Office of Science. SDSS-IV acknowledges support and resources from the Centre for High-Performance Computing at the University of Utah. The SDSS web site is {\tt www.sdss.org}.

The participating Institution in SDSS-II include the American Museum of Natural History, Astrophysical Institute Potsdam, University of Basel, University of Cambridge, Case Western Reserve University, University of Chicago, Drexel University, Fermilab, the Institute for Advanced Study, the Japan Participation Group, Johns Hopkins University, the Joint Institute for Nuclear Astrophysics, the Kavli Institute for Particle Astrophysics and Cosmology, the Korean Scientist Group, the Chinese Academy of Sciences (LAMOST), Los Alamos National Laboratory, Max-Planck-Institut f\"ur Astronomie (MPIA Heidelberg), Max-Planck-Institut f\"ur Astrophysik (MPA Garching), New Mexico State University, Ohio State University, University of Pittsburgh, University of Portsmouth, Princeton University, the United States Naval Observatory, and the University of Washington.

SDSS-IV is managed by the Astrophysical Research Consortium for the Participating Institutions of the SDSS Collaboration including the  Brazilian Participation Group, the Carnegie Institution for Science, Carnegie Mellon University, the Chilean Participation Group, the French Participation Group, Harvard-Smithsonian Center for Astrophysics, Instituto de Astrof\'isica de Canarias, The Johns Hopkins University, Kavli Institute for the Physics and Mathematics of the Universe (IPMU) / University of Tokyo, Lawrence Berkeley National Laboratory, Leibniz Institut f\"ur Astrophysik Potsdam (AIP),  Max-Planck-Institut f\"ur Astronomie (MPIA Heidelberg), Max-Planck-Institut f\"ur Astrophysik (MPA Garching), Max-Planck-Institut f\"ur Extraterrestrische Physik (MPE), National Astronomical Observatory of China, New Mexico State University, New York University, University of Notre Dame, Observat\'ario Nacional / MCTI, The Ohio State University, Pennsylvania State University, Shanghai Astronomical Observatory, United Kingdom Participation Group, Universidad Nacional Aut\'onoma de M\'exico, University of Arizona, University of Colorado Boulder, University of Oxford, University of Portsmouth, University of Utah, University of Virginia, University of Washington, University of Wisconsin, Vanderbilt University, and Yale University.

All data taken as part of SDSS-IV is scheduled to be released to the community in fully reduced form at regular intervals through dedicated data releases. The MaNGA data used in this work is part of SDSS data release 13 \citep{SDSS_DR13}, publicly available at {\tt http://www.sdss.org/dr13/manga/manga-data/}.
\end{small}


\bibliography{bib14}
\bibliographystyle{mnras}

%
%
\noindent \hrulefill

\noindent $^1$ University of Cambridge, Cavendish Astrophysics, Cambridge, CB3 0HE, UK.
\\$^2$ University of Cambridge, Kavli Institute for Cosmology, Cambridge, CB3 0HE, UK.
\\$^{3}$ Institute of Cosmology and Gravitation, University of Portsmouth, Dennis Sciama Building, Portsmouth, PO1 3FX, UK.
\\$^4$ European Southern Observatory, Karl-Schwarzchild-str., 2, Garching b. Munchen, 85748, Germany.
\\$^{5}$ Universit\'e Lyon 1, Observatoire de Lyon, Centre de Recherche Astrophysique de Lyon and Ecole Normale
Sup\'erieure de Lyon, 9 avenue Charles Andr\'e, Saint-Genis Laval, F-69230, France.
\\$^6$ Department of Astronomy, University of Winsconsin-Madison, 475 N. Charter Street, Madison, WI 53706-1582, USA.
\\$^{7}$ South East Physics Network (SEPNet), www.sepnet.ac.uk
\\$^{8}$ Sternberg Astronomical Institute, Moscow State University, Moscow, Russia.
\\$^{9}$ Apache Point Observatory and New Mexico State University, PO Box 59, Sunspot, NM 88349-0059, USA.
\\$^{10}$ Universidad de Antofagasta, Unidad de Astronomía, Avenida Angamos 601, Antofagasta, 1270300, Chile.
\\$^{11}$ Department of Physics and Astronomy, University of Utah, 115 S. 1400 E., Salt Lake City, UT 84112, USA.
\\$^{12}$ Kavli Institute for the Physics and Mathematics of the Universe (WPI), The University of Tokyo Institutes for Advanced Study, The University of Tokyo, Kashiwa, Chiba 277-8583, Japan.
\\$^{13}$ Department of Physics and Astronomy, Bates College, Lewiston, ME 04240, USA.
\\$^{14}$ McDonald Observatory, The University of Texas at Austin, 2515 Speedway, Stop C1402, Austin, TX 78712, USA.
\\$^{15}$ Center for Astrophysical Sciences, Department of Physics and Astronomy, The Johns Hopkins University, Baltimore, MD 21218, USA.
\\$^{16}$ Space Telescope Science Institute, 3700 San Martin Drive, Baltimore, MD 21218, USA.
\\$^{17}$ Departamento de Fisica, Facultad de Ciencias, Universidad de La Serena, Cisternas 1200, La Serena, Chile.
\\$^{18}$ School of Physics and Astronomy, University of St. Andrews, North Haugh, St. Andrews, KY16 9SS, UK.
\\$^{19}$ Department of Physics and Astronomy, University of Kentucky, 505 Rose Street, Lexington, KY 40506-0055, USA.


\appendix

\section{Computing the star formation rate}
\label{AppA}

In this work we compute the SFR using the extinction-corrected H$\alpha$ flux from the MaNGA data. We calculate the reddening from the Balmer decrement, by using the $\rm H \alpha  / H \beta$ ratio and a \cite{Calzetti2001a} attenuation curve with $\rm R_V = 4.05$. The theoretical value for the Balmer line ratio is taken from \cite{Osterbrock2006}, assuming case B recombination ($ \mathrm{H \alpha  / H \beta=2.87 }$). We note that the use of extinction curve of \cite{Cardelli1989} (or the modification by \citealt{O'Donnell1994}) with $\rm R_V=3.1$ yields very similar results for the 3600 $\rm \AA$ to 7000 $\rm \AA$ wavelength range considered in this work. 

The simplest approach to computing SFR would be to exclude regions (spaxels) which are not classified as star forming by making use of a hard cut on a BPT diagnostic diagram. We consider this approach partly unsatisfactory because of the rather arbitrary nature of the demarcation lines. Instead we suggest a framework where the total H$\alpha$ flux is the combination of a contribution from H\textsc{ii} regions, related to star formation, and a contribution from old evolved stars (LIERs) and LIER-like diffuse ionised gas. In order to disentangle the two components we rely on their different lines ratios, following a similar approach to \cite{Blanc2009}. In this work we choose to use [SII]/H$\alpha$, which is the most sensitive line ratio to the presence of LIER emission studied in \textit{Paper I}. Defining $\rm f_{SF}$ to be the fraction of H$\alpha$ flux due to star formation and $\rm f_{L}$ the fraction of H$\alpha$ flux due to LIERs, we have 

\begin{equation}
\rm f_{SF}+ f_{L}=1, \\ 
\left( \frac{[SII]}{H\alpha} \right)= f_{SF} \left( \frac{[SII]}{H\alpha} \right)_{SF}+ f_{L} \left( \frac{[SII]}{H\alpha} \right)_{L}
\end{equation}
where $\rm \left( \frac{[SII]}{H\alpha} \right)_{SF}=0.4$ and $\rm \left( \frac{[SII]}{H\alpha} \right)_{L}=0.9$ are the characteristic line ratios for star forming regions and LIERs, respectively, whose numerical values are obtained in \textit{Paper I}, as the representative line ratio values for SF and eLIER galaxies respectively. To calculate the SFR we solve for $\rm f_{SF}$ for each spaxel and then sum the total H$\alpha$ luminosity due to star formation within the FoV of each MaNGA galaxy. We convert the H$\alpha$ luminosity to SFR using the conversion factor from \cite{Kennicutt1998}. We note that this approach leads to very similar numerical results to a hard cut on SF spaxels based on a BPT demarcation line. For example, using the same demarcation line as \textit{Paper I} on the [SII] BPT diagram to classify SF spaxels and only using those spaxels in the SFR calculation leads to SFR which strongly correlate with our estimate for purely SF galaxies (Pearson correlation coefficient of 0.997). The SFR calculated using the LIER correction is, on average, systematically larger by only 0.03 dex from that computed using only SF spaxels with a average scatter of 0.07 dex. 

\begin{figure} 
\label{AppA1}
\includegraphics[width=0.5\textwidth, trim=100 60 100 80, clip]{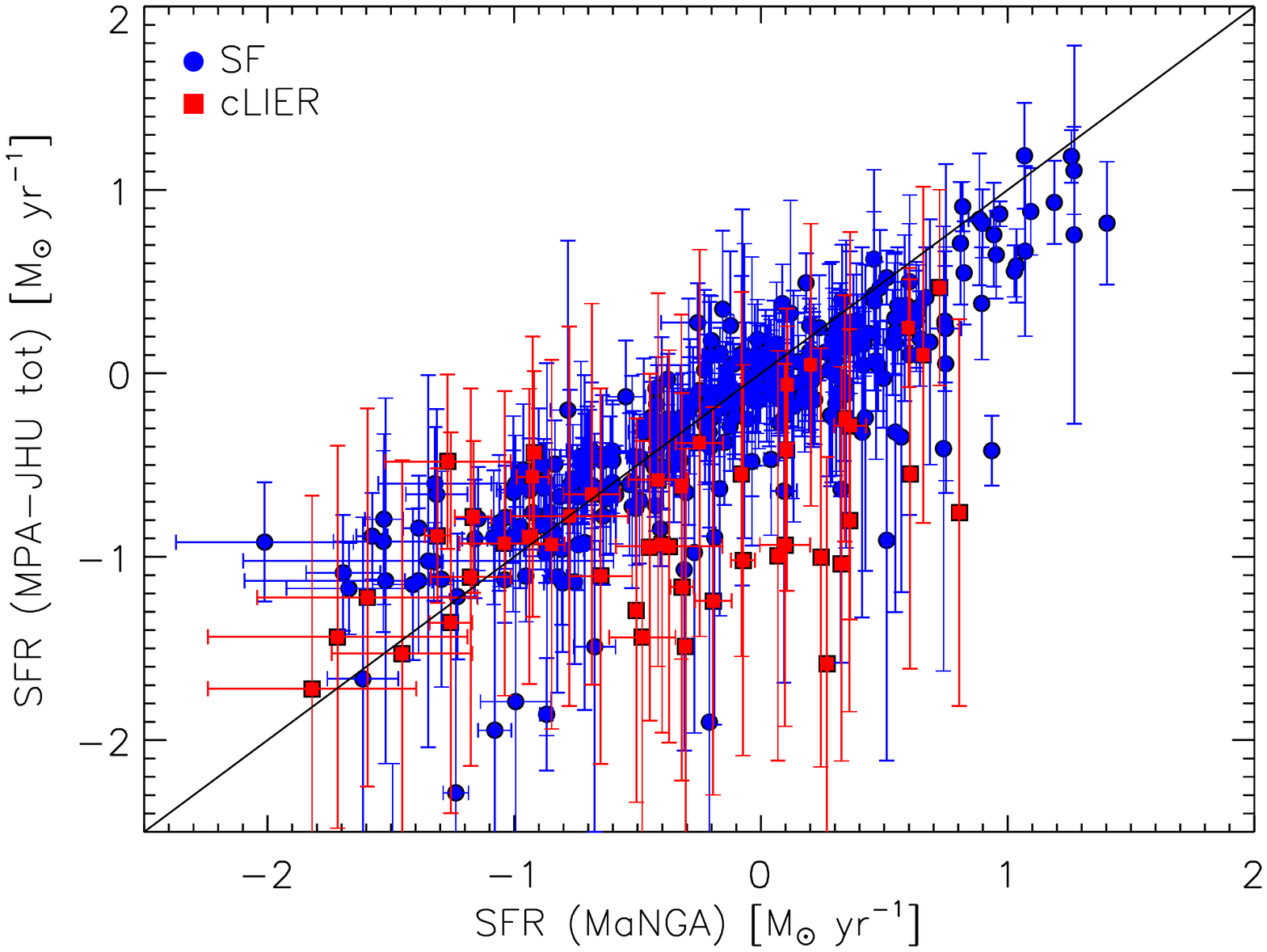}
\caption{The relation between the SFR estimates from the MPA-JHU catalogue \protect\citep{Brinchmann2004} based on legacy SDSS spectroscopy and MaNGA (as calculated in this work) for SF and cLIER galaxies. The one-to-one relation is highlighted with a black solid line.}

\label{AppA2}
\end{figure}

Our SFR estimates are also in excellent agreement with the SFR reported in the MPA-JHU catalogue \citep{Brinchmann2004}, based on legacy SDSS spectroscopy (Pearson correlation coefficient of 0.8, dispersion of 0.4 dex). As shown in Fig. \ref{AppA2}, the offset from the one-to-one relation is substantially larger for cLIERs (0.5 dex) than SF galaxies (0.07 dex), which is not surprising given the poor leverage of the SDSS 3$''$ fibre in calculating the SFR in cLIERs. Further details on the SFR estimators for MaNGA galaxies and aperture corrections is beyond the scope of this section and will be presented in future work.

\section{Additional material}

We present here SDSS g-r-i composite images and H$\alpha$ flux maps all the cLIER and eLIER galaxies used in this work.

\begin{table*}
\caption{Identifiers and fundamental properties of the galaxy sample used in this work. Table for the full sample available online.}

\begin{tabular}{ l  c c c c c r c c c c}

\hline 
\hline 
MaNGA ID	 & 	NSAID 	&	 RA 			& 	Dec 		 &  z	&  Plate  	& IFU design & $\rm \log (M_\star) $	& $\rm log(SFR)$	& $\mathrm{R_{e}}$ 		& EM. class			\\
	       		&			& 	J2000 deg  	&  	J2000 deg	& 	&		&		     &	$\rm [M_\odot]$		&$\rm [M_\odot~yr^{-1}$]	& [$''$] 			& 		\\
\hline 
 1-23929    &   25589 & 259.174713 &   56.927792 & 0.028 & 7991    & 9101   & 10.086  &  -0.353  &   8.768   &   SF  \\   
  1-24006    &   25673 & 257.833832 &  56.991253 	&0.031  &7991  &   6103   &  9.734  &  -0.356    & 5.565   &   SF   \\  
 1-24099     &  25775 & 258.027618   & 57.504009 	&0.028  &7991   &  1902   & 10.462  &   NaN   &   3.624  & eLIER   \\  
 1-24124     &  25801 & 258.530273  &  57.477432 	&0.027 &7991    & 3703    & 9.768    &  NaN  &  4.818   &   LL   \\  
 1-24354    &   26050 & 261.083344   & 56.876629 &0.028 & 7991    & 6102     &9.900     & NaN    & 6.634     & LL    \\ 
 1-24368      & 26065  &261.321045   & 56.951878 &0.025  &7991    &12704    &10.042  &  -0.317  &  12.865    &  SF   \\  
 1-25819     &  27618 & 259.874573   & 55.432957 &0.072  &7991    & 3701    &11.069  &  < -1.256    & 3.073     & LL   \\ 
 1-25911     &  27716  & 260.346985   & 55.876526 &0.140 & 7991   &  9102  &  11.471 & NaN  &   6.755   &   LL    \\ 
 1-37965     &  40459  & 48.642723    & 0.532375 &0.021  &8082     &9101     &9.432    &-1.614   &  8.656     & SF     \\
 1-37973     &  40468  & 49.108551   &  0.321643 &0.021  &8082    & 3702    & 9.258    &-1.311    & 5.266   & cLIER    \\ 
 ... & & & & & & & & & & \\
\hline
\end{tabular}
\begin{flushleft}
\small
\textit{Column notes:} 
(1) MaNGA ID is the primary identifier denoting a unique galaxy within the MaNGA survey; (2) NSA ID is the identification number of the MaNGA galaxy target from the NSA parent catalogue; (3-5) taken from NSA; (6-7) The combination of plate and IFU design uniquely identifies each MaNGA observation (equivalent to the plate-fibre-MJD spectral identifiers from legacy SDSS); (8) $\rm M_\star $ from fit to the integrated photometry from the NSA; (9) SFR from MaNGA data, as computed in this work; (10) Half-light radius from a S\'ersic fit to the r-band legacy SDSS photometry, from the NSA; (11) Emission line class as defined in this work: SF - star forming, cLIER - central LIER, eLIER - extended LIER, LL - line-less.
\end{flushleft}

\label{tableA1}
\end{table*}

\begin{table*}
\caption{Photometric and morphological properties for the galaxy sample used in this work. Table for the full sample available online.}

\begin{tabular}{ l  c c c c c c c c c}

\hline 
\hline 
MaNGA ID	 & b/a S\'ersic	 &  n S\'ersic	 &  $\phi$ S\'ersic & $\rm R_{90}/R_{50}$ & 	 GZ $\rm p_{disc}$ & $\rm \sigma_\star (0.5 Re)$	& $NUV - r $		& $u-r$			&$g-r$  	\\
	       		 &			& 	 		& [deg]	 	    & 				     & 					     &$\rm [ km~s^{-1}]$	 	&mag			&mag 			& mag 	 \\
\hline 
 1-23929         &   0.366  &   1.008   & 12.278  &   2.389    & 0.420    &55.170    &   3.260    & 1.986    & 0.651 \\
 1-24006          &  0.259  &   0.829 &  177.225 &    2.532   &  0.774   & 53.330   &   3.108    & 1.754   &  0.579\\
 1-24099       & 0.666   &  3.893   &127.189  &   3.029    & 0.077   &123.840   &  5.497    & 2.344    & 0.738\\
 1-24124        & 0.446   &  1.100   & 37.710   &  2.351     &0.113   & 56.480    & 4.748   &  2.085   &  0.625\\
 1-24354        &   0.288   &  0.616  &  61.907   &  2.007    & 0.978    &65.840    & 5.422  &   2.391    & 0.731\\
 1-24368         &  0.725   &  1.904  & 173.668  &   2.949   &  0.530  &  44.620   &   2.419    & 1.612  &   0.543\\
 1-25819        &   0.586   &  6.000  & 101.348  &   3.279   &  0.237   &361.190   &  5.707   &  2.832   &  0.853\\
 1-25911        &  0.787    & 5.115   & 88.744    & 3.384    & 0.549   &226.980   & 6.369   &  2.367 &    0.577\\
 1-37965        &  0.901   &  1.153  & 137.144  &   3.110   &  0.981    &29.830   &  2.671   &  1.406   &  0.449\\
 1-37973        &  0.251  &   1.412  & 146.287   &  2.545   &  0.896    &43.310  &   2.397   &  1.693   &  0.617\\
  ... & & & & & & & & &\\
\hline
\end{tabular}

\begin{flushleft}
\small
\textit{Column notes:} 
(1) MaNGA ID, same as above; (2) Major to minor axis ratio of the S\'ersic fit to the legacy SDSS r-band, from the NSA; (3) S\'ersic index of the fit, from the NSA; (4) Position angle of the S\'ersic fit, from the NSA; (5) Concentration parameter ($R_{90}$ and $R_{50}$ are Petrosian radii containing respectively 90\% and 50\% of the r-band light), from the NSA; (6) De-biased probability of the galaxy containing a disc, from Galaxy Zoo 2 \protect\cite{Willett2013}; (7) Average stellar velocity dispersion within 0.5 $\rm R_e$, from MaNGA data; (8-10) $NUV-r$, $u-r$ and $g-r$ colours from the NSA.

\end{flushleft}

\label{tableA2}
\end{table*}

\begin{figure*} 
\includegraphics[width=0.8\textwidth, trim=30 200 0 150, clip]{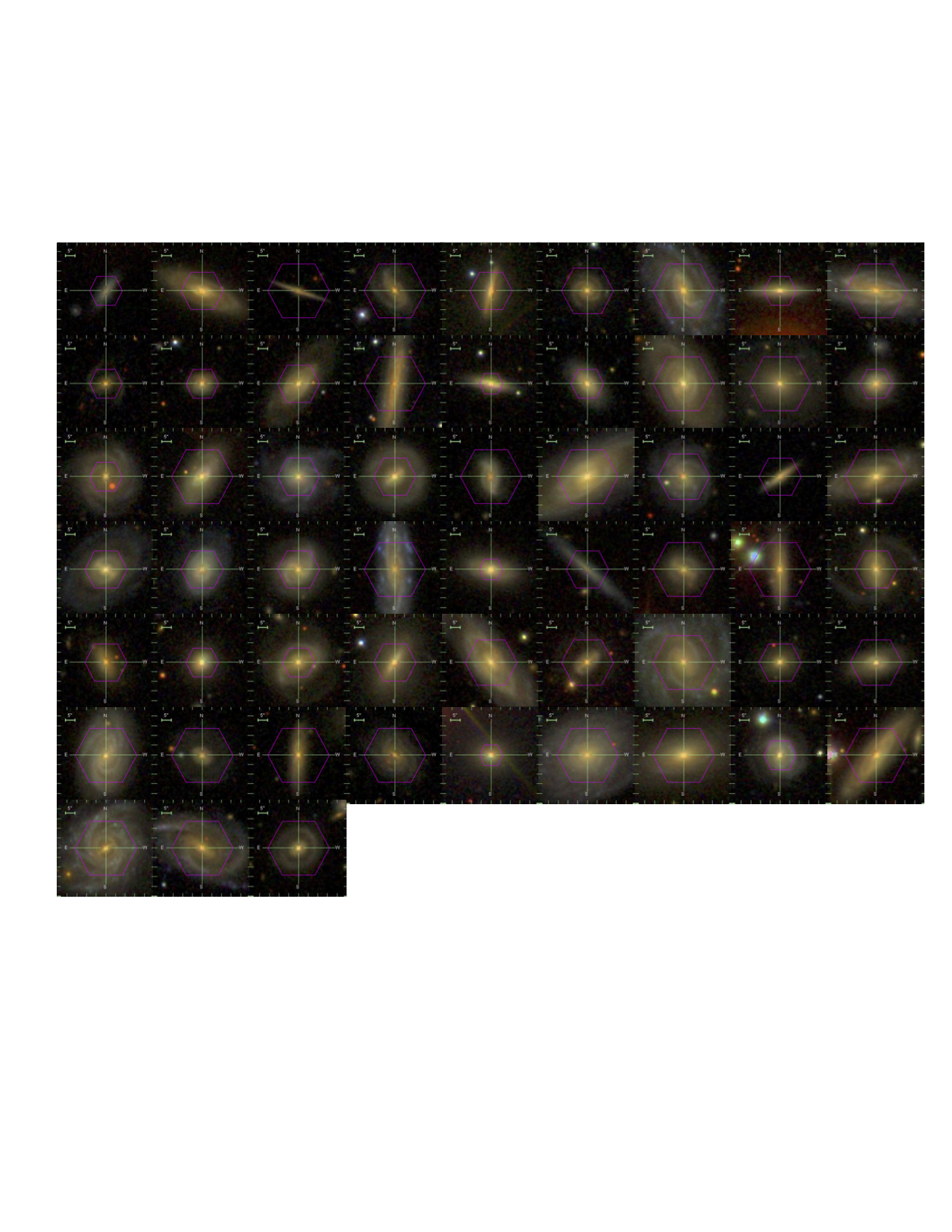}
\includegraphics[width=0.8\textwidth, trim=30 200 0 150, clip]{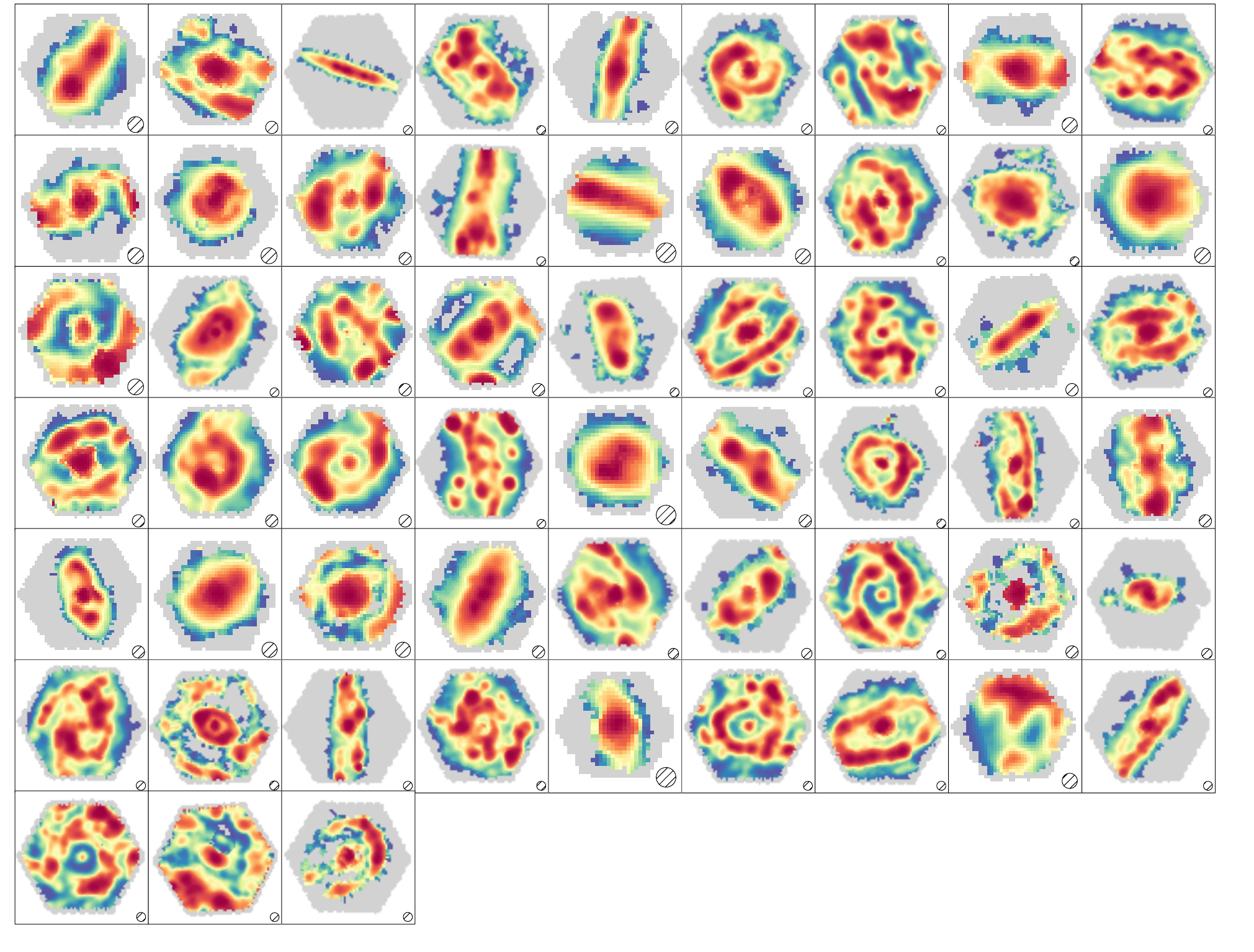}
\caption{\textit{Top}: SDSS g-r-i colour composite images (with MaNGA IFU bundle FoV superimposed) for the sample of cLIER galaxies studied in this work. \textit{Bottom}: MaNGA H$\alpha$ flux maps for the sample of cLIER galaxies studied in this work.}
\label{figAPP1}
\end{figure*}

\begin{figure*} 
\includegraphics[width=0.8\textwidth, trim=30 200 0 150, clip]{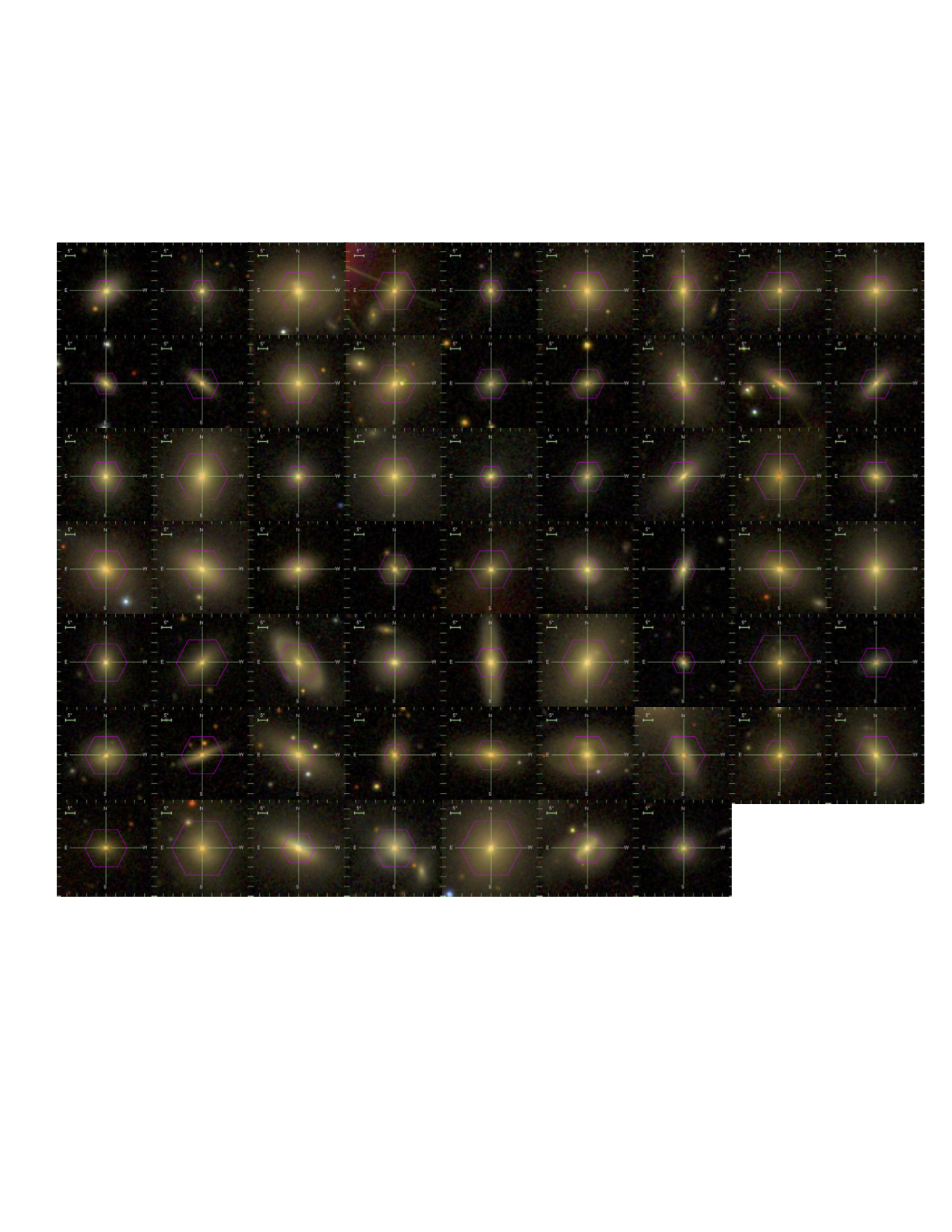}
\includegraphics[width=0.8\textwidth, trim=30 200 0 150, clip]{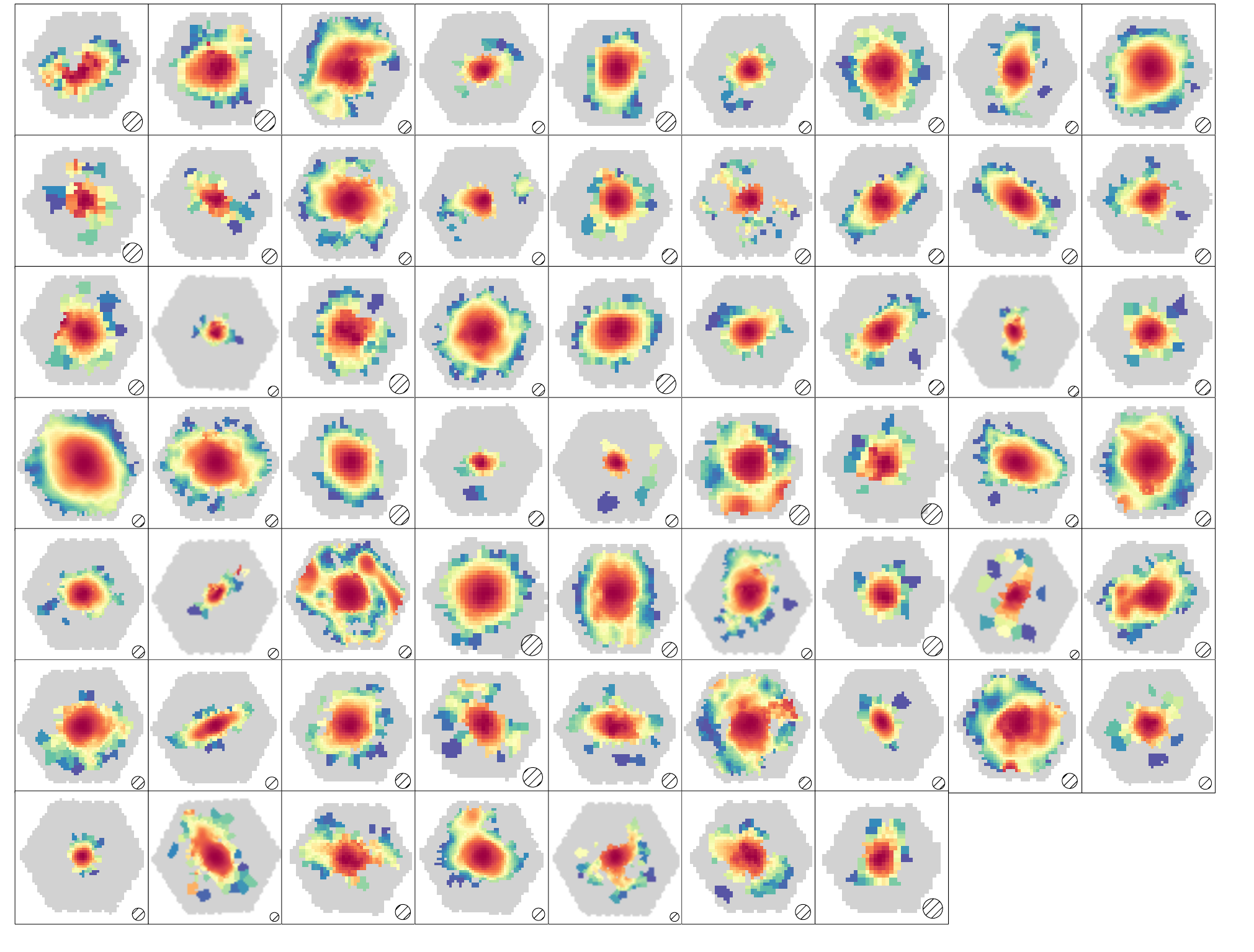}
\caption{\textit{Top}: SDSS g-r-i colour composite images (with MaNGA IFU bundle FoV superimposed) for the sample of eLIER galaxies studied in this work. \textit{Bottom}: MaNGA H$\alpha$ flux maps for the sample of eLIER galaxies studied in this work.}
\label{figAPP1}
\end{figure*}

\bsp
\label{lastpage}
\end{document}